\documentclass{JHEP3}
\usepackage{graphicx}
\usepackage{latexsym}

\def\a{\alpha}\def\b{\beta}\def\g{\gamma}\def\d{\delta}\def\e{\epsilon}

\def\l{\lambda}

\def\PRL #1 #2 #3{{\em Phys. Rev. Lett. \/} {\bf#1} (#2) #3}
\def\NPB #1 #2 #3{{\em Nucl. Phys. \/} {\bf B#1} (#2) #3}
\def\NPBFS #1 #2 #3 #4{{\em Nucl. Phys. \/} {\bf B#2} [FS#1] (#3) #4}
\def\CMP #1 #2 #3{{\em Commun. Math. Phys. \/} {\bf #1} (#2) #3}
\def\PRD #1 #2 #3{{\em Phys. Rev. \/} {\bf D#1} (#2) #3}
\def\PLA #1 #2 #3{{\em Phys. Lett. \/} {\bf #1A} (#2) #3}
\def\PLB #1 #2 #3{{\em Phys. Lett. \/} {\bf B#1} (#2) #3}
\def\JMP #1 #2 #3{{\em J. Math. Phys. \/} {\bf #1} (#2) #3}
\def\PTP #1 #2 #3{{\em Prog. Theor. Phys. \/} {\bf #1} (#2) #3}
\def\SPTP #1 #2 #3{{\em Suppl. Prog. Theor. Phys. \/} {\bf #1} (#2) #3}
\def\AoP #1 #2 #3{{\em Ann. of Phys. \/} {\bf #1} (#2) #3}
\def\PNAS #1 #2 #3{{\em Proc. Natl. Acad. Sci. USA} {\bf #1} (#2) #3}
\def\RMP #1 #2 #3{{\em Rev. Mod. Phys. \/} {\bf #1} (#2) #3}
\def\PR #1 #2 #3{{\em Phys. Reports \/} {\bf #1} (#2) #3}
\def\AoM #1 #2 #3{{\em Ann. of Math. \/} {\bf #1} (#2) #3}
\def\UMN #1 #2 #3{{\em Usp. Mat. Nauk \/} {\bf #1} (#2) #3}
\def\FAP #1 #2 #3{{\em Funkt. Anal. Prilozheniya \/} {\bf #1} (#2) #3}
\def\FAaIA #1 #2 #3{{\em Functional Analysis and Its Application \/} {\bf
#1} (#2) #3}
\def\BAMS #1 #2 #3{{\em Bull. Am. Math. Soc. \/} {\bf #1} (#2)
#3} \def\TAMS #1 #2 #3{{\em Trans. Am. Math. Soc. \/} {\bf #1} (#2)
#3}
\def\InvM #1 #2 #3{{\em Invent. Math. \/} {\bf #1} (#2) #3}
\def\LMP #1 #2 #3{{\em Letters in Math. Phys. \/} {\bf #1} (#2) #3}
\def\IJMPA #1 #2 #3{{\em Int. J. Mod. Phys. \/} {\bf A#1} (#2) #3}
\def\AdM #1 #2 #3{{\em Advances in Math. \/} {\bf #1} (#2) #3}
\def\RMaP #1 #2 #3{{\em Reports on Math. Phys. \/} {\bf #1} (#2) #3}
\def\IJM #1 #2 #3{{\em Ill. J. Math. \/} {\bf #1} (#2) #3}
\def\APP #1 #2 #3{{\em Acta Phys. Polon. \/} {\bf #1} (#2) #3}
\def\TMP #1 #2 #3{{\em Theor. Mat. Phys. \/} {\bf #1} (#2) #3}
\def\JPA #1 #2 #3{{\em J. Physics \/} {\bf A#1} (#2) #3}
\def\JSM #1 #2 #3{{\em J. Soviet Math. \/} {\bf #1} (#2) #3}
\def\MPLA #1 #2 #3{{\em Mod. Phys. Lett. \/} {\bf A #1} (#2) #3}
\def\JETP #1 #2 #3{{\em Sov. Phys. JETP \/} {\bf #1} (#2) #3}
\def\JETPL #1 #2 #3{{\em  Sov. Phys. JETP Lett. \/} {\bf #1} (#2) #3}
\def\PHSA #1 #2 #3{{\em Physica} {\bf A#1} (#2) #3}
\def\CQG #1 #2 #3{{\em Class. Quantum Grav. \/} {\bf #1} (#2) #3}
\def\SJNP #1 #2 #3{{\em Sov. J. Nucl. Phys. (Yadern.Fiz.) \/} {\bf #1} (#2) #3}

\def\be{\begin{equation}}
\def\ee{\end{equation}}
\def\ba{\begin{array}} \def\ea{\end{array}}
\def\bea{\begin{eqnarray}}
\def\eea{\end{eqnarray}}

\title {Dynamics of Higher Spin Fields and Tensorial Space}

\author{{I. Bandos$^{a,d}$, X. Bekaert$^{b}$, J.~A.~de Azc\'arraga$^{a}$,
D. Sorokin$^{c}$ and M. Tsulaia$^{c,e}\footnote{On leave from
Bogoliubov Laboratory of Theoretical Physics, JINR, 141980, Dubna,
Russia }$
\\
\\
  {\small $^{a}$ \it Departamento de F\'{\i}sica Te\'orica and IFIC (CSIC-UVEG),
 46100-Burjassot (Valencia), Spain}
\\
{\small $^{b}$ \it Institut des Hautes \'Etudes Scientifiques,}
\\
{\small \it Le Bois-Marie, 35 route de Chartres, 91440
Bures-sur-Yvette, France}
\\
{\small $^{c}$ \it INFN Sezione di Padova ${\&}$ Dipartimento di
Fisica ``Galileo Galilei",}
\\
{\small  \it Universit\`{a} degli Studi di Padova, 35131 Padova,
Italy}
\\
{\small $^{d}$ \it Institute for Theoretical Physics, NSC KIPT,
61108 Kharkov, Ukraine}
\\
{\small $^{e}$ \it Institute of Physics, GAS 380077 Tbilisi,
Georgia}
 }}

\abstract{The structure and the dynamics of massless higher spin
fields in various dimensions are reviewed with   an emphasis on
conformally invariant higher spin fields. We show that in
$D=3,4,6$ and $10$ dimensional space--time the conformal higher
spin fields constitute the quantum spectrum of a twistor--like
particle propagating in tensorial spaces of corresponding
dimensions. We give a detailed analysis of the field equations of
the model and establish their relation with known formulations of
free higher spin field theory.}


\preprint{DFPD 05/TH/03, IHES/P/05/06}

\begin{document}

\section{Introduction}

The twistor--like particles propagating in tensorial (super)spaces
put forward in \cite{Bandos:1998vz,Bandos:1999qf} have the
interesting property of being related to massless higher--spin
fields. A key point in this construction is the extension of the
conventional $D$--dimensional space--time, parametrized by
coordinates $x^m$, with extra directions parametrized by
antisymmetric tensor coordinates $y^{mn\cdots q}$. In these models
the tensorial coordinates correspond to the helicity degrees of
freedom of the quantum states of the system in ordinary space--time.

In particular, the $Sp(8,\mathbb R)$--invariant twistor
superparticle model produces upon quantization an infinite tower
of higher-spin fields in $D=4$ space--time, where each and every
massless representation of the conformal group $SU(2,2)\subset
Sp(8,\mathbb R)$ appears only once. Thus this model turns out to
be a realization of a Kaluza--Klein--like mechanism conjectured by
Fronsdal \cite{fronsdal1}. Upon  performing an appropriate Fourier
transform in twistor space and integrating out the extra tensorial
variables $y^{mn}$ one finds that
\cite{Vasiliev:2001zy,Misha,Plyushchay:2003gv,Plyushchay:2003tj}
the wave functions of these higher-spin fields satisfy the
unfolded higher spin field equations, which are known to be an
appropriate framework in which a self--consistent interaction of
higher spin fields can be introduced (see \cite{V01} for
references and \cite{Misha03-04b} for recent progress).

A (bosonic) tensorial space is parametrized by symmetric $n\times
n$ matrix coordinates $X^{\a\b}=X^{\b\a}$ ($\a,\b=1,\ldots, n$)
linearly transformed by the group $GL(n,\mathbb R)$. The dimension
of such a space is ${n(n+1)}\over 2$. For appropriate {\sl even}
values of $n$ the link with ordinary $D$--dimensional space--time
coordinates $X^m$ is made by decomposing $X^{\a\b}$ in a basis of
symmetric $n\times n$ gamma--matrices
\begin{equation}X^{\a\b}=
x^m\g_m^{\a\b}+y^{mn\cdots q}\g_{mn\cdots q}^{\a\b}\,,\quad
(m=0,1,\ldots,D-1;\quad \a,\b=1,\cdots,n)\,, \label{xab}
\end{equation}
with completely antisymmetric $y^{mn\cdots q}$.

 Physically interesting examples of
 tensorial spaces are
\begin{itemize}
 \item
$n=2$. The  tensorial  space has  dimension 3 and corresponds to a
conventional $D=3$
 space--time without extra $y$--coordinates;
\item
$n=4$. The tensorial space is 10--dimensional and corresponds to
$D=4$ space--time enlarged with 6 extra coordinates $y^{mn}$;
\item
$n=8$. The tensorial space has  dimension 36. It is parametrized by the
coordinates $x^m$ of $D=6$ space--time and by the $3{1\over
2}{6\choose 3}=30$ components of an $SO(3)$ triplet of
anti--self--dual coordinates $y^{mnp}_I$ (I=1,2,3) (where
${n\choose k}={n! \over{(n-k)!\cdot k!}}$);
\item
$n=16$. The tensorial space has dimension 136. It is parametrized by the
coordinates $x^m$ of $D=10$ space--time and by ${1\over
2}{{10}\choose 5}=126$ anti--self--dual coordinates $y^{mnlpq}$.
\end{itemize}
For all these cases the space--time dimension $D$ is related to the
(spinor) dimension $n$ by the formula $n=2(D-2)$.

One can also consider tensorial spaces with Grassmann directions
parametrized by coordinates $\theta_i^\alpha$ ($i=1,\ldots,N;\,
\a,\b=1,\cdots,n$), thus dealing with tensorial superspaces (as
$\Sigma^{({{n(n+1)}\over 2}|n)}$ for $N=1$ \cite{Bandos:2003us}).
A physically interesting example is provided by the supergroup
manifolds $OSp(N|n)$.

The $n=2$ case is well known and corresponds to conventional field
theories in $D=3$ space--time. The physical $D=4$ space--time
higher spin contents of the $n=4$ model has been studied in detail
in
\cite{Bandos:1999qf,Vasiliev:2001zy,Misha,Plyushchay:2003gv,Plyushchay:2003tj},
while only generic properties of higher dimensional
generalizations of these models and their generalized
(super)conformal structure have been discussed
\cite{Bandos:1999qf,blps,Vasiliev:2001zy,othertopics,Misha,Plyushchay:2003gv,Gelfond:2003vh,V01}.
Though the $n=8$ and 16 tensorial superparticles were quantized in
\cite{Bandos:1999qf}, no detailed analysis of the corresponding
spectra of higher spin fields and of their equations of motion in
$D=6$ and 10 space--time, which should follow from the tensorial
field equations, has been carried out so far. So the main purpose
of this paper is to consider in detail the physical $D=6$ and
$D=10$ space--time contents of the $n=8$ and $n=16$ tensorial
models. We will show that, analogously to the $n=4$, $D=4$ case, the
first quantization of the tensorial particle produces a
representation of $Sp(2n,\mathbb R)$ which decomposes into an
infinite sum of irreducible representations (irreps) of the
conformal group $Spin(2,D)\subset Sp(2n,\mathbb R)$. In addition to
scalar and spinor fields the infinite sets of $D=6$ and
$D=10$ fields associated with these representations consist of
massless higher spin fields of mixed symmetry whose field
strengths (or curvatures)\footnote{In what follows we shall freely
use either the name `field strength' or `curvature'.} are
self--dual.

We shall show that the space--time wave equations for the
conformal higher spin fields in $D=4,6$ and 10 obtained from a
scalar and a spinor field equation in the corresponding $n=2(D-2)$
tensorial space are `geometric' in the sense that they are written
in terms of gauge invariant linearized curvature tensors. In $D=4$
these higher spin field curvatures are a straightforward
generalization of the linearized Riemann curvature and satisfy the
same cyclic first and second Bianchi identities as the latter. The
higher spin `Riemann' curvatures are related to the generalized
curvatures introduced by de Wit and Freedman \cite{deWit:1979pe}
via an appropriate (anti)symmetrization of indices.  In $D=4$
Minkowski space--time the free higher spin field equations have
been known for a long time, since the paper by Dirac \cite{Dirac}.
In the form presently known as `Bargmann--Wigner equations' they
were analyzed from a group--theoretical point of view in
\cite{Bargmann}. In the massless case these are the `geometric'
equations for higher spin curvatures\footnote{For a pedagogical
review, see Chapter 1 and Sec. 6.9 of \cite{Buchb}, and
\cite{Corson} for general relativistic $D=4$ wave equations and
historical references.}. As was shown in \cite{Bracken:1982ny},
the Bargmann--Wigner form of the equations for the massless $D=4$
higher spin fields is the most convenient one to exhibit their
conformal invariance. Geometric free field equations were written
by Vasiliev \cite{LV,Misha03-04b} in a moving frame--like (or
vielbein--like) formulation for completely symmetric gauge fields
propagating in $AdS_D$ space--time. In a metric--like formulation
geometric free field equations for arbitrary higher spin gauge
fields in flat space--time of any dimension were proposed in
\cite{Siegel:1986zi,BB1}. These are a generalization of the $D=4$
Bargmann--Wigner equations for the higher spin field curvatures.
All these equations have a common drawback, namely, that for
higher spin fields they cannot be directly obtained from an action
principle. Note that the free action for arbitrary higher spin
fields constructed in \cite{Siegel:1986zi} is quadratic in
derivatives of the integer spin field potentials and is of the
first order in the derivatives of the half--integer field
potentials. In the case of symmetric tensor fields it reproduces
the Fronsdal \cite{F} and Fang--Fronsdal \cite{FF} actions (see
\cite{fields} for more details). Such actions give rise to the
second-- or first--order differential equations for the higher
spin field {\it potentials} which (except for the spins $s=1,3/2$
and 2) cannot be directly rewritten in terms of higher spin {\it
curvatures} because of the following reason.

A {\it spin--$s$ curvature} (or {\it field strength}) is obtained
by taking $[s]$ curls of the corresponding {\it gauge field
potential} (the bracket denotes integer part). Thus, by
definition, a local free higher spin field equation formulated in
terms of the field strength must contain at least $[s]$ partial
derivatives. As a result, for $s>2$  the local geometric higher
spin equations contain more than two derivatives and if they were
Lagrangian the corresponding actions would be of $[s]$-th order in
derivatives. So naively the geometric formulations of higher spin
theories seem to suffer from the higher--derivative problem which
states that free theories whose physical degrees of freedom obey
differential equations of order strictly greater than two for
bosons, and one for fermions, have ghosts \cite{Pais}. However,
the higher spin fields circumvent this problem in a rather subtle
way: the higher spin field strength equations reduce to second or
first order differential equations for the corresponding integer
or half integer higher spin potentials.

As we have already mentioned, local second and first order
differential equations for massless bosonic and fermionic higher
spin fields described by symmetric (spinor)--tensors and
corresponding actions were constructed in \cite{F,FF} and for
generic higher spin fields in \cite{Siegel:1986zi,fields}. In such
formulations the higher spin gauge fields and the gauge parameters
satisfy algebraic (trace) constraints. These restrictions on the
higher spin gauge fields and parameters look unnatural and,
basically, two ways of removing them have been proposed (see
\cite{d} for recent reviews of problems of higher spin field
theory).

One way is to renounce the locality of the theory. Non--local
actions for unconstrained higher-spin gauge fields leading to
non--local geometric field equations were constructed by Francia
and Sagnotti \cite{Francia:2002aa} and generalized to mixed
symmetry fields in \cite{dMH}.

The second way of relaxing the trace constraints, keeping locality
at the same time, is by introducing a new field called `compensator'
\cite{Francia:2002pt,Sagnotti:2003qa}. The resulting field equations
are non--Lagrangian in the sense that using only the proper higher
spin gauge fields and the compensator one is not able to construct
an action from which these equations follow. In order to construct a
Lagrangian  one
has to introduce extra auxiliary fields
\cite{Pashnev:1998ti,Sagnotti:2003qa}.
 The number of these auxiliary fields increases with the
value of the spin of the `basic' field.

The generalized cohomologies introduced in \cite{Olver:1983,DuboisV}
(and extended to mixed symmetry fields in \cite{BB1}) further
clarified the geometrical structure of higher spin gauge theories.
We will see that, for example, the unconstrained gauge invariance and
the Bianchi identities of the spin $s$ field strength (with
$s>{1\over 2}$) are elegantly summarized in terms of the generalized
nilpotency of an exterior derivative $\partial$,
$\partial^{[s]+1}=0$ that leads one to the introduction of a
generalized cohomology\footnote{To avoid confusion with the
standard exterior derivative $d$ ($d^2=0$), the exterior derivative
obeying the higher order nilpotency  condition is denoted by
$\partial$. }. The cohomological results of
\cite{Olver:1983,DuboisV} are crucial for the possibility of
relating the geometric curvature equations to the compensator
equations via a generalization of the spin 3 Damour--Deser identity
\cite{DD}. The latter establishes the relationship between the trace
of the spin 3 field curvature with a curl of the kinetic operator
acting on the spin 3 field potential in the Fronsdal formulation
\cite{F}. For bosons with $s>2$ this identity expresses
$(s-2)$ curls of the spin $s$ Fronsdal kinetic operator as the
trace of the field strength \cite{DD,BB2}. For fermions with
$s>3/2$, it expresses $(s-3/2)$ curls of the spin $s$ Fang--Fronsdal
kinetic operator as the gamma--trace of the field strength. The
explicit relationship between the higher spin curvature equations
and the compensator equations was shown in \cite{BB2} for integer
spin fields and will be extended to half integer spin fields in this
paper. This clarifies how local geometric equations of order $[s]$
are equivalent to (Fang)-Fronsdal equations, and thus explains
how higher spin gauge fields circumvent the higher--derivative problem.

The paper is organized as follows.  The geometric formulation of
higher spin field theory in terms of the generalized curvatures
and its application to the description of conformally invariant
higher spin fields are discussed in Sec. \ref{conformalHS}. In
Sec. \ref{Preonicequ} we briefly review the twistor--like particle
model in tensorial space with a generic $n$. Sec. \ref{Spectrum}
is devoted to the study of the quantum spectrum of the particle in
a tensorial space with $n=2(D-2)$ and $D=3,4,6$ and $10$. It is
shown that this spectrum consists of an (infinite) set of
$D=3,4,6$ and $10$ massless (higher spin gauge) fields obeying
geometric equations on their curvatures invariant under the
generalized conformal group $OSp(1|4(D-2))\supset SO(2,D)$. The
relation of these equations to those of Francia and Sagnotti, and
of Fang and Fronsdal generalized to mixed symmetry fields is
demonstrated. For clarity, we first review the well known $D=3$
and $D=4$ cases in Subsections 4.1 and 4.2. Then in Sec. 4.3 we
pass to the simpler $D=10$ case, and devote Sec. 4.4 to the
technically more involved $D=6$ case demonstrating that in both of
them the spectrum consists of massless self--dual higher spin
fields whose field strengths satisfy geometric equations.
In the Conclusions we discuss several possible directions for
future research.

\setcounter{equation}0
\section{Massless higher spin fields in any dimension}\label{conformalHS}

We review here the general properties and equations which (spinor)
tensor fields in $D$--dimensional space--time should obey to
describe massless higher spin states associated with an appropriate
unitary irreducible representation of the Poincar\'e group.
Group--theoretical arguments and the quantum consistency of the
theory require that the massless higher spin fields   be gauge
fields and that their gauge invariant field strengths (or
curvatures)   satisfy irreducibility conditions which constitute the
geometrical higher spin field equations.

\subsection{Geometric equations}

In order to describe a massless unitary irreducible representation
of the Poincar\'e group \linebreak \hbox{$ISpin(1,D-1)$}, a
(spinor--)tensor field strength should
\begin{itemize}
  \item[(i)] be \textit{irreducible} with respect to $GL(D)$. Though, it
  should be noted that when the spinor structure is introduced,
  and/or the field strength satisfies a tracelessness
  condition (which requires the introduction of a metric),
  the general linear group $GL(D)$ is restricted to its
  $SO(1,D-1)$ subgroup or to $Spin(1,D-1)$ covering of the latter.

Strictly speaking, the conventional notion of spin (or helicity) is
only well defined in $D\leq 4$. In $D> 4$ we will loosely call the
spin the number $s$ that characterizes a massless irreducible
representation of $ISpin(1,D-1)$ which corresponds to a Young
diagram  with $[s]$ columns (where $[s]$ denotes the integer part of
$s$). As will become clear in a moment, a spin $s$ field strength
should be characterized by the Young diagram\footnote{ Young
diagrams will be denoted by
$(\ell_1,\ell_2,\ldots,\ell_c)$. They consist of a finite number
$c$ of rows with decreasing numbers of boxes $\ell_1\geqslant
\ell_2\geqslant \ldots\geqslant \ell_c > 0$.} $([s],[s] , r_1
,\ldots , r_{c-2})$
\begin{equation} \label{YD}
\matrix{ & \quad {}^{\overbrace{\hspace{3.8cm}}^{s}} \quad 
\cr c &  \left\{ \matrix{\cr\cr {}_{\begin{picture}(120,25)(0,5)
{\linethickness{.250mm
\put(00,00){\line(1,0){50}}
\put(00,10){\line(1,0){80}}
\put(00,20){\line(1,0){80}}
\put(00,35){\shortstack{...}}\put(15,35){\shortstack{...}}
\put(30,35){\shortstack{...}}
\put(00,25){\shortstack{...}}\put(15,25){\shortstack{...}}
\put(30,25){\shortstack{...}}
\put(00,40){\line(1,0){110}}
\put(00,50){\line(1,0){110}}
\put(00,60){\line(1,0){110}}
\put(00,00){\line(0,1){60}}
\put(10,00.0){\line(0,1){10}} \put(20,00.0){\line(0,1){10}}
\put(27,02.0){\shortstack{...}}
\put(40,00.0) {\line(0,1){10}} \put(50,00.0){\line(0,1){10}}
\put(10,10.0){\line(0,1){10}} \put(20,10.0){\line(0,1){10}} 
\put(27,12.0){\shortstack{...}}
\put(40,10.0){\line(0,1){10}} \put(50,10.0){\line(0,1){10}}
\put(58,12.0) {\shortstack{...}}
\put(70,10.0){\line(0,1){10}} \put(80,10.0){\line(0,1){10}}}
\put(10,40.0){\line(0,1){20}} \put(20,40.0){\line(0,1){20}}
\put(30,40.0){\line(0,1){20}} \put(40,40.0){\line(0,1){20}}
\put(50,45){\shortstack{...}}\put(70,45){\shortstack{...}}
\put(50,55){\shortstack{...}}\put(70,55){\shortstack{...}}
\put(90,40.0){\line(0,1){20}} \put(100,40.0){\line(0,1){20}}
\put(110,40.0){\line(0,1){20}}
}%
\end{picture}} \cr\cr }\right. \cr &
\cr & \quad {\underbrace{\hspace{1.6cm}}_{r_{c-2} \leq s}}
\hspace{2.6cm} }  \equiv   ([s],[s] , r_1 ,\ldots , r_{c-2})
\end{equation}
the first two rows of which have an equal length
$\ell_1=\ell_2=[s]$ (the other rows are arbitrary and $c$ is the
length of the first column or   number of rows). The field strength
tensor is expressed in the antisymmetric basis, in the sense that
each column corresponds to a set of antisymmetric indices, {\it
e.g.} the electromagnetic field strength $F_{mn}$ (s=1) corresponds
to the Young diagram $(1,1)$.

Note, however, that setting apart the conformal fields of spin $s$
that are characterized by {\it rectangular} diagrams (Sec. 2.2),
the label $s$ does {\it not} fully determine the field. Thus, for
non--conformal fields the full set of labels in (\ref{YD}) is
required.


  \item[(ii)] be \textit{harmonic}, in the sense that
it is closed and co--closed (transversal) with respect to each
{\it set} of antisymmetric indices.

If a field strength is closed, then the generalized Poincar\'e
lemmas of \cite{Olver:1983,DuboisV,BB1} imply that the field
strength is locally \textit{exact}, {\it i.e.} it is equal to
$[s]$ curls of a corresponding gauge field potential characterized
by the Young diagram $([s],r_1,\ldots,r_{c-2})$ obtained by
removing the first row of the field strength diagram. Indeed, the
first row of the field strength is made of $[s]$ partial
derivatives of the gauge field potential, the symmetry properties
making them act as a curl on a given set of antisymmetric indices.

  \item[(iii)] obey the \textit{Dirac} equation if $s$ is half--integer.

  \item[(iv)] be \textit{traceless i.e}.
$\gamma$-traceless in the case of half--integer $s$  and traceless
with respect to any pair of tensorial indices.
\end{itemize}

\noindent Conditions (i), (ii) and (iv) were proposed as  field
equations for arbitrary bosonic mixed symmetry fields in
\cite{BB1} as a generalization of the $D=4$ Bargmann--Wigner
equations \cite{Bargmann}. These conditions are not completely
independent. For instance, transversality is a consequence of
(iii) and of $\g$--tracelessness in the fermionic case. In the
bosonic case, transversality follows from (i), the closedness
condition and tracelessness. Conditions (i) and (iv) insure that
the on--shell field strength is irreducible with respect to
$Spin(1,D-1)$. In other words, the linearized curvature is equal
to its Weyl part. Moreover, it can be shown that a tensor obeying
(i)-(iv) corresponds to a unitary irreducible representation of the
compact subgroup $Spin(D-2)$ (SO(2) for $D=4$) of the ``little
group" of $Spin(1,D-1)$. This representation  corresponds to a Young
diagram $([s],r_1,\ldots,r_{c-2})$ with up to $(D-3)$ rows, so that
$c\leq D-2$. In turn, this irrep induces a massless unitary irrep
of the Poincar\'e group $ISpin(1,D-1)$ characterized by a {\it
discrete} spin\footnote{Note that, as in
$D=4$, by restricting the little group to its compact subgroup we
remove from the consideration the unphysical massless `continuous'
spin representations.}. This generalization of the $D=4$
Bargmann--Wigner  results provides a rigorous proof
\cite{BB3}\footnote{For a simple explicit example see Sec. 2 and
Appendix A of \cite{BBC}.} of the fact that the previously
proposed equations of motion of a generic mixed symmetry field in
flat space--time \cite{Labastida:1986ft,Labastida:1986zb,BB1,dMH}
describe the proper physical higher spin degrees of freedom. Mixed
symmetry fields on $AdS$ have been considered in
\cite{Brink:2000ag} where the situation is more complicated.

Let us note that mixed symmetry fields appear in various physical
models. For instance, massive mixed symmetry fields are part of
the spectrum of first--quantized strings. One can also obtain
mixed symmetry fields by dualizing completely symmetric gauge
fields in a space--time of dimension $D>4$ (see for instance
\cite{Hull:2001iu,BB1,BB2,Boulanger:2003vs}). The problem of
constructing free field actions for the mixed symmetry fields is
efficiently solved in the framework of the BRST approach
\cite{Burdik:2001hj}.\footnote{Originally inspired by string field
theory,  the BRST approach has been used for the analysis of the
higher spin spectra of string states in the tensionless limit
\cite{Bonelli:2003kh,Sagnotti:2003qa,Bekaert:2003uc}. Recently, a
first--order `parent' field theory was constructed along the lines
of the BRST approach in the context of higher--spin gauge theories
\cite{Barnich:2004cr}.} Alternatively, `multiform' and `hyperform'
calculus proved to be efficient mathematical tools to deal with
the theory of mixed symmetry fields
\cite{Olver:1983,DuboisV,BB1,dMH}.

After this group--theoretical introduction let us make contact
with the unconstrained non--local formulation of
\cite{Francia:2002aa,Francia:2002pt,Sagnotti:2003qa}. The first
two conditions (i)--(ii) allow us to express the field strength in
terms of a gauge field potential. The field strength is
automatically invariant under gauge transformations with
unconstrained gauge parameters. Now we have to distinguish among
integer spin fields (bosons) and half--integer spin fields
(fermions).

For {\it bosons}, upon solving for (i) and (ii) in terms of the
gauge potential the remaining condition on the field strength is
its tracelessness. The trace of the field strength with respect to
indices belonging to the first two columns is equal to $s-2$ curls
of the Labastida kinetic operator \cite{Labastida:1986ft} which
generalizes the Fronsdal kinetic operator to mixed symmetry
fields. The tracelessness of the field strength thus states that
the Labastida kinetic operator is $\partial^{s-2}$ closed. Then
the generalized Poincar\'e lemma implies that (since for the spin
$s$ fields $\partial^{s+1}=0$, and $s+1-(s-2)=3$) the Labastida
kinetic operator is $\partial^3$--exact (for $s>2$) in the sense
that it is equal to a sum of compensator fields differentiated
three times. Each of these compensators corresponds to a Young
diagram obtained by removing from the diagram
$(s,r_1,\ldots,r_{c-2})$ three cells in different columns. In the
case of the symmetric higher spin fields only one compensator
field appears, and the resulting form of the higher spin equations
coincides with that of the compensator equations of
\cite{Francia:2002pt,Sagnotti:2003qa}. This relation between the
geometric equations on the higher spin curvature and the
compensator equations was demonstrated in \cite{BB2,BB3} for an
arbitrary bosonic mixed symmetry field. The compensator fields can
be gauged away by fixing the traces of the gauge parameters.
Alternatively, they can be expressed in a non--local way in terms
of the kinetic operators thereby producing non--local equations of
\cite{Francia:2002aa,dMH}.

In the {\it fermionic} case of the half integer spin fields, the
relationship between the higher derivative geometric equations and
the first order differential equations for the spinor--tensor
gauge field potentials has not been considered in the literature
before, and here we fill this gap. The reasoning follows the same
lines as in the case of the integer spin fields. One can
explicitly check that the $\g$--trace of the field strength with
respect to the indices belonging to the first two columns is equal
to $s-3/2$ curls of the Labastida fermionic kinetic operator
\cite{Labastida:1986zb} which generalizes the Fang--Fronsdal
kinetic operator to  mixed symmetry tensor--spinor fields. The
$\g$--tracelessness of the field strength which plays the role of
the geometric generalization of the  Dirac equation thus states that
the Labastida kinetic operator is $\partial^{s-3/2}$--closed. Then
(since for half integer spin $s$ fields $\partial^{s+1/2}=0$, and
$s+{1\over 2}-(s-{3\over 2})=2$) the generalized Poincar\'e lemma
implies that  the Labastida kinetic operator is
$\partial^2$--exact (for $s>3/2$) in the sense that it is equal to a sum of
fermionic compensator fields differentiated twice.
 Each of these compensators corresponds to
a Young diagram obtained by removing from the diagram
$(s-1/2,r_1,\ldots,r_{c-2})$ two cells in different columns. In
the case of the symmetric tensor--spinor fields only one
compensator field appears, and the resulting form of the higher
spin equations coincides with that of the compensator equations of
\cite{Sagnotti:2003qa}. The compensator fields can be gauged away
by fixing the $\g$--traces of the gauge parameters. Alternatively,
they can be expressed in a non--local way in terms of the kinetic
operators thereby producing the non--local equations of
\cite{Francia:2002aa,dMH}.

We shall now consider in more detail how these general
considerations work in the case of conformally invariant higher
spin fields.

\subsection{Conformal higher--spin fields in various dimensions}
Among the massless (free) higher spin fields there is an interesting
and important subclass of fields whose (at least linearized)
equations of motion are conformally invariant. Note, however, that
in (anti) de Sitter spaces also partially massless higher spin
fields can be conformally invariant (see \cite{dw} and references
therein).

 In $D\geqslant 3$ the necessary and
sufficient condition for a unitary irreducible representation of
the Poincar\'e group $ISpin(1,D-1)$ to be extendable to a unitary
irrep of the conformal group $Spin(2,D)$ is that the
representation is induced from the restriction of the `massless
little group' to its compact subgroup $Spin(D-2)$, and corresponds
to a rectangular Young diagram with columns of length equal to
$D/2$ \cite{Siegel:1988gd}. Physically speaking, this means that
free conformal unitary field theories are described by massless
fields with discrete helicity whose field strength is
\textit{chiral}, {\it i.e.} self--dual or anti--self--dual. For
the odd space--time dimensions this leaves the scalar field and
the Dirac fermion as the only possibilities\footnote{Conformal
higher spin field theories in $D=3$, considered {\it e.g.} in
\cite{cd3}, are of a higher--derivative Chern--Simons type and
hence do not have propagating physical degrees of freedom.}.  For
even space--time dimensions $D=2c$, the spinorial index should be
Weyl and the field should be self--dual (or anti--self--dual) with
respect to each set of antisymmetric indices (with the same
chirality for all of them) \footnote{In space--times of Lorentz
signature with double even dimensions, {\it i.e.} $c=2k,~D=4k$,
the chiral field strengths are complex. }. This leads to the field
strengths with symmetry properties characterized by   {\it
rectangular} Young diagrams $[s]\times D/2$ with columns of equal
length $c=D/2$ and rows of equal length $[s]$ \footnote{More
general classes of $D$--dimensional conformal fields and equations
which include those corresponding to non--unitary field theories
have been discussed in \cite{Shaynkman:2004vu}.}.

Such self--dual mixed symmetry fields appear in the $(4,0)$
conformal theory in six dimensions, which has been conjectured to
be an analogue of the (2,0) conformal theory living on the
M5-brane worldvolume \cite{Hull:2000rr}.
 Infinite sets of
self--dual higher spin fields also appear in the $D=6$ and $10$
spectrum of states of a first--quantized particle propagating in
 tensorial spaces with $n=2(D-2)$, their equations of motion
being invariant under the generalized superconformal group
$OSp(1|4(D-2))\supset SO(2,D)$. This will be the subject of
Secs. \ref{Preonicequ} and \ref{Spectrum}, while below we shall
consider the geometrical formulation of the free $D$--dimensional
theory of the self--dual higher spin fields characterized by the
rectangular Young tableaux $[s]\times D/2$ along the general lines
discussed in the previous Subsection.

To simplify a bit notation let us introduce a cumulative index
\begin{equation}\label{D/2}
\left[{D\over 2}\right]:= [m_1\cdots m_{D\over 2}],
\end{equation}
which stands for $D\over 2$ antisymmetrized indices. Wherever
several cumulative indices appear, they will denote the
corresponding groups of antisymmetrized indices, {\it e.g.}
\begin{equation}\label{D/2,12}
\left[{D\over 2}-1\right]_1\left[{D\over 2}\right]_2:= [m_1\cdots
m_{{D\over 2}-1}],[n_1\cdots n_{D\over 2}]\,.
\end{equation}
Two cumulative indices which denote the same number of
antisymmetric indices are assumed to be symmetric, {\it  e.g.}
\begin{equation}\label{D/2,121}
\left[{D\over 2}\right]_1\left[{D\over 2}\right]_2=\left[{D\over
2} \right]_2\left[{D\over 2}\right]_1\,.
\end{equation}
Whenever it is unavoidable, we shall use conventional and
cumulative indices together.

To illustrate our notation let us recall the familiar example of
four--dimensional linearized gravity. In this case $D=4$ and
$s=2$, and the Riemann tensor is denoted as
\begin{equation}\label{ri}
R_{m_1m_2,\,n_1n_2}(x)=-R_{m_2m_1,n_1n_2}(x)=R_{n_1n_2,\,m_1m_2}(x)
\equiv R_{_{\left[2\right]_1\,\left[2\right]_2}}(x)
=R_{_{\left[{2}\right]_2\,\left[2\right]_1}}(x)\,.
\end{equation}
The cyclic Bianchi identity and the differential Bianchi identity,
respectively, imply that
\begin{equation}\label{b1}
R_{[m_1m_2,\,n_1]n_2}\equiv
R_{_{\left[{3}\right]_1\,\left[1\right]_2}}=0\,,
\end{equation}
\begin{equation}\label{b2}
\partial_{[m_3}\,R_{ m_1m_2],\,n_1n_2}\equiv
\partial_1\,R_{_{\left[{2}\right]_1\,\left[2\right]_2}}=0\,,
\end{equation}
where the subscript of $\partial_1$ means that the exterior
derivative is antisymmetrized together with the first group
$\left[{2}\right]_1=[m_1m_2]$ of  antisymmetric indices.

\subsubsection{Integer spin fields}

 With this notation in mind let
\begin{equation}\label{Rcurv}
R_{_{\left[{D\over 2}\right]_1\cdots\,\left[{D\over 2}\right]_s}}=
R_{m_1\cdots m_{_{D\over 2}},\cdots,\, q_1\cdots \,q_{_{D\over
2}}}
\end{equation}
 be the curvature (or the field strength) of a conformal
integer spin $s$ characterized by a   rectangular Young diagram with
${D\over 2}$ rows and $s$ columns.

The requirement (i) of Sec. \ref{conformalHS} then implies that
the curvature tensor is symmetric under exchange of any two
cumulative indices and that it satisfies the cyclic Bianchi
identity
\begin{equation}\label{i}
R_{_{\left[{D\over 2}+1\right]_1\,\left[{D\over 2}-1\right]_2
\,\left[{D\over 2}\right]_3\cdots\left[{D\over
2}\right]_s}}=0.\nonumber
\end{equation}
The curvature is closed if it satisfies the differential Bianchi
identity
\begin{equation}
\partial_{[m_1}R_{m_2\cdots
n_{{D\over 2}+1}],\,{\left[{D\over 2}\right]_2\cdots\left[{D\over
2}\right]_s}}=0\,, \label{closed}
\end{equation}
and is co--closed
(or transverse) if
\begin{equation}\label{trans}
\partial^n R_{n\left[{D\over 2}-1\right]_1\,,\,
\left[{D\over 2}\right]_2\cdots\left[{D\over
2}\right]_s }=0\,.
\end{equation}

As in the case of $D=4$ gravity, the Bianchi identity
(\ref{closed}) can be written as an exterior derivative acting as
a curl on one of the groups of antisymmetric indices of the
multiform $R_{_{\left[{D\over 2}\right]_1\cdots\left[{D\over
2}\right]_s}}$
\begin{equation}\label{multi}
\partial_1\,R_{_{\left[{D\over 2}\right]_1\cdots\left[{D\over
2}\right]_s}}=0\,,
\end{equation}
where the subscript of $\partial_1$ means that the exterior
derivative index is antisymmetrized together with the first group
$\left[{D\over 2}\right]_1$ of the antisymmetric indices.

 Let us denote in general by
\begin{equation}\label{di}
\partial_i\equiv
1\,\otimes\,\cdots\,\otimes\,\partial_{m_i}\,\otimes\,\cdots\,\otimes
1 \quad ( i=1,\cdots, s )
\end{equation}
 the exterior derivative (curl)
antisymmetrized with the $i$-th group $\left[{D\over 2}\right]_i$
of antisymmetric indices \footnote{Actually, if we worked with
differential {\it multiforms}, the  differential operator
(\ref{di}) would correspond to the operator $
1\,\otimes\,\cdots\,\otimes\,d\,\otimes\,\cdots\,\otimes 1\equiv
(1\,\otimes\,\cdots\,\otimes\,dx^{m_i}\,\otimes\,\cdots\,\otimes
1)\,
\partial_{m_i}\,, $ where the exterior derivative $d$ stands in the $i$-th
place and acts on the $i$-th block of the {\it multiform}
characterized by the cumulative antisymmetric tensor index
$[{D\over 2}]_i$.
 However, since we would like to keep track of the indices,
we prefer to use the definition (\ref{di}) where the partial
derivative $\partial_{m_i}$ acts as a curl within the $i$--th
cumulative index.}. Then,
$$
\partial_i\,\partial_j \equiv
1\,\otimes\,\cdots\,\otimes\,\partial_{m_i}\,\otimes\,\cdots\,\otimes\,
\partial_{m_j}\,\otimes
\cdots\,\otimes 1=\partial_j\,\partial_i, \quad (\partial_i)^2=0
$$
are ``curled" with $\left[{D\over 2}\right]_i$
and $\left[{D\over 2}\right]_j$, {\it etc.}

Let us now introduce the differential operator
\begin{equation}\label{partial}
\partial := \sum^{s}_{i=1}\,\partial_i\,.
\end{equation}
In view of the nilpotency of the exterior derivative
($\partial_i^2=0$ for each $i$) the differential operator
$\partial$ satisfies the higher order nilpotency condition
\begin{equation}\label{partials}
\partial^{s+1} = \partial\,\partial^s=0\,,
\quad {\rm where}\quad  \partial^s:=s\,
\prod_{i=1}^s\,\partial_i=s\,\partial_1\otimes\partial_2
\otimes\cdots\otimes\partial_s\,.
\end{equation}

According to the generalized Poincar\'e lemma of \cite{DuboisV},
the Bianchi identity (\ref{closed}) implies that (at least
locally) the curvature is the $s$-th derivative of a potential,
which in the `multiform' notation reads
\begin{equation}
R_{_{\left[{D\over 2}\right]_1\cdots\left[{D\over
2}\right]_s}}=\partial_1 \cdots \partial_s \,
\varphi_{_{\left[{D\over 2}-1\right]_1\cdots\left[{D\over
2}-1\right]_s}}\,,\label{fielstr}
\end{equation}
where, as defined in (\ref{di}), each $\partial_i$ acts as an
exterior derivative (curl) on the corresponding group
$\left[{D\over 2}-1\right]_i$ of antisymmetric indices. Using the
notation (\ref{partials}), eq. (\ref{fielstr}) can be written in a
more schematic way as follows
$$
R={1\over s}\,\partial^s\,\varphi\,,
$$
which is the generalization to the spin $s$ fields of the well
known expression of the electromagnetic field strength in terms of
the curl of the spin 1 field potential $F=\partial\,A$.

The field $\varphi\equiv \varphi_{_{\left[{D\over
2}-1\right]_1\cdots\left[{D\over 2}-1\right]_s}}$ is the conformal
gauge field potential of integer spin $s$ characterized by the
rectangular Young diagram $s\times({D\over 2}-1)$, so it is
symmetric under the exchange of any two of the $s$ cumulative
indices $[{D\over 2}-1]_i$ and satisfies a cyclic Bianchi identity
similar to (\ref{i}),
$$\varphi_{_{_{\left[{D\over 2}\right]_1\,\left[{D\over
2}-2\right]_2 \,\left[{D\over 2}-1\right]_3\cdots\left[{D\over
2}-1\right]_s}}}=0.
$$

Let us note that de Wit and Freedman \cite{deWit:1979pe}
constructed curvature tensors out of the $s$ derivatives of the
symmetric higher spin gauge fields $\varphi_{m_1\cdots\,m_s}(x)$
in an alternative way. Their curvatures have two groups of $s$
symmetric indices and they are symmetric or antisymmetric under
the exchange of these groups of indices depending on whether $s$
is even or odd
\begin{equation}\label{dwf}
R_{m_1\cdots m_s,\,n_1\cdots n_s}=(-1)^s\,R_{n_1\cdots
n_s,\,m_1\cdots m_s}.
\end{equation}
For $D=4$, the tensor (\ref{dwf}) is related to  the tensor
(\ref{fielstr})
  by the antisymmetrization of each pair $[m_i,\,n_i]$ of
indices of the former. In what follows we will work with the
generalized Riemann curvatures.

Due to (\ref{partials}), the field strengths (\ref{fielstr}) and
(\ref{dwf}) are invariant under the following gauge
transformations of the gauge potential \cite{DuboisV}
\begin{eqnarray}
\delta \varphi_{_{\left[{D\over
2}-1\right]_1\cdots\,\left[{{D\over
2}}-1\right]_s}}&=&\partial_{1}\,
 \xi_{_{\left[{D\over
 2}-2\right]_1\,\left[{D\over 2}-
1\right]_2\cdots\,\left[{{D\over 2}}-1\right]_s}}
 +\partial_{2}\,
\xi_{_{\left[{D\over 2}-1\right]_1\,\left[{D\over 2}-2\right]_2\cdots
\, \left[{D\over 2}-1\right]_s}} + \cdots
\nonumber\\
& =& \sum_{i=1}^{s}\,\partial_{i}\, \xi_{_{\left[{D\over 2}-1\right]_1\,\cdots
 \,\left[{D\over 2}-2\right]_i\cdots\,\left[{D\over
 2}-1\right]_s}}\,,
 \label{gaugetransf}
\end{eqnarray}
where $\xi(x)$ is an  unconstrained gauge function characterized
by the Young diagram $(s,\ldots,s,s-1)$ with $[{D\over 2}-1]$
rows.

When the conditions (i) a (ii) of Sec. \ref{conformalHS} on the
integer spin curvature are resolved in terms of the gauge
potential, the only one which remains is (iv), {\it i.e.}
tracelessness of the curvature tensor in any pair of its indices
belonging to different cumulative indices
\begin{equation}\label{tracel}
tr\,R_{_{\left[{D\over 2}\right]_1\cdots\left[{D\over
2}\right]_s}}=0\,.
\end{equation}
This is the field equation that generalizes the linearized
Einstein equation $R^p_{~m,\,pn}=R_{mn}=0$ for spin 2.

Recall that for conformal fields in even--dimensional
space--times, the self--duality condition
$$
R_{_{\left[{D\over 2}\right]_1\cdots\left[{D\over 2}\right]_s}}=
\pm\frac{i^{^{{D\over 2} +1}}}{{D\over 2}
!}\,\,\epsilon_{_{\left[{D\over 2}\right]_1,\,n_1\cdots
n_{_{D\over 2}}} }\,R^{n_1\cdots n_{_{D\over
2}}\,\,}{}_{_{\left[{D\over 2}\right]_2\cdots\,\left[{D\over
2}\right]_s}}$$
 is the actual
field equation because tracelessness and transversality follow
from the self--duality condition provided the curvature satisfies
the Bianchi identities.

Analyzing the form of the left hand side of eq. (\ref{tracel}) in
terms of the gauge field potential (\ref{fielstr}) one gets the
generalization of the spin 3 Damour--Deser identity \cite{DD}
\begin{equation}
tr\,R_{_{\left[{D\over 2}\right]_1\cdots\left[{D\over
2}\right]_s}} =\partial_1\cdots \partial_{s-2}\,
{G}_{_{\left[{D\over 2}-1\right]_1\cdots\left[{D\over
2}-1\right]_{s-2}\,\left[{D\over 2}-1\right]_{s-1}\,\left[{D\over
2}-1\right]_s}}\label{DDgen}
\end{equation}
where $G$ is the kinetic operator acting on the gauge field
potential \cite{Labastida:1986ft}
\begin{eqnarray}
{G}_{_{\left[{D\over 2}-1\right]_1\cdots\,\left[{D\over
2}-1\right]_s}}& =&\square\,{\varphi}_{_{\left[{D\over
2}-1\right]_1\,\cdots\,\left[{D\over
2}-1\right]_s}}-\sum_{i=1}^s\partial_i\,\partial^m\varphi_{_{\left[{D\over
2}-1\right]_1\cdots\,,\,m\left[{D\over
2}-2\right]_i\,,\cdots\,\left[{D\over 2}-1\right]_{s}}}\nonumber\\
&&+ \sum_{j>i=1}^s\partial_{i}\,\partial_{j}\,\eta^{mn}\,
\varphi_{_{\left[{D\over 2}-1\right]_1\cdots\,,m\left[{D\over
2}-2\right]_{i}\,,\cdots\,,n\left[{D\over
2}-2\right]_{j}\,,\cdots\,\left[{D\over 2}-1\right]_s}}\,,
\end{eqnarray}
where (in accordance with our notation and convention) the sums
are taken over the terms with the exterior derivative $\partial_i$
indices antisymmetrized with those of the corresponding group
$\left[{D\over 2}-2\right]_{i}$.

When the curvature tensor satisfies the tracelessness condition
(\ref{tracel}) the left hand side of eq. (\ref{DDgen}) vanishes,
which implies that the multiform $G$ is $\partial^{s-2}$--closed.
In virtue of the generalized Poincar\'e lemma \cite{DuboisV} this
means that (at least locally) $G$ is $\partial^3$--exact, {\it
i.e.} has the form \cite{BB2}
\begin{equation}
{G}_{_{\left[{D\over 2}-1\right]_1\cdots\,\left[{D\over
2}-1\right]_s}}=\sum_{k>j>i=1}^s\,\partial_i\,\partial_j\,\partial_k\,
\rho_{_{\left[{D\over 2}-1\right]_1\cdots\,,\left[{D\over
2}-2\right]_{i}\,,\cdots\,,\left[{D\over
2}-2\right]_{j}\,,\cdots\,\left[{D\over
2}-2\right]_{k}\,,\cdots\,\left[{D\over 2}-1\right]_s}}\,,
\label{comp}
\end{equation}
where the tensor field $\rho(x)$ is characterized by the Young
diagram $(s,\ldots,s,s-3)$ with $[{D\over 2}-1]$ rows. The tensor
$\rho(x)$ is called `compensator' field since its gauge
transformation compensates the non--invariance of the kinetic
operator $G(x)$ under the unconstrained local variations
(\ref{gaugetransf}) of the gauge field potential $\varphi(x)$.
Therefore, eq. (\ref{comp})  generalizes to arbitrary rectangular
Young diagrams  the compensator equation given in
\cite{Francia:2002pt,Sagnotti:2003qa}.

The gauge variation of $G(x)$ is
\begin{equation}\label{GG}
\d{G}_{_{\left[{D\over 2}-1\right]_1\cdots\,\left[{D\over
2}-1\right]_s}} =
\sum_{k>j>i=1}^s\,\partial_i\,\partial_j\,\partial_k\,
\eta^{mn}\,\xi_{_{\left[{D\over
2}-1\right]_1\cdots\,,\left[{D\over
2}-2\right]_{i}\,,\cdots\,,\left[{D\over 2}-2\right]_{j}\,
m,\cdots\,\left[{D\over 2}-2\right]_{k}\,n,\cdots\,\left[{D\over
2}-1\right]_s}}\,
\end{equation}
and it is compensated by the gauge shift of the field $\rho(x)$
with the trace of the gauge parameter
\begin{equation}\label{rhog}
\delta\rho_{_{\left[{D\over 2}-2\right]_{1}\,\left[{D\over
2}-2\right]_{2}\,\left[{D\over 2}-2\right]_{3}\, \left[{D\over
2}-1\right]_4\,\cdots\,\left[{D\over
2}-1\right]_s}}=\eta^{mn}\,\xi_{_{\left[{D\over
2}-2\right]_{1}\,m\,,\,\left[{D\over
2}-2\right]_{2}\,n,\,\left[{D\over 2}-2\right]_{3}\, \left[{D\over
2}-1\right]_4\,\cdots\,\left[{D\over 2}-1\right]_s}}\,.
\end{equation}
So the compensator can be gauged away by choosing a gauge
parameter $\xi(x)$ with the appropriate trace. Then the equations
of motion of the gauge field $\phi(x)$ become the second order
differential equations of Labastida, which generalize those of
Fronsdal for mixed symmetry fields
\begin{eqnarray}
{G}_{_{\left[{D\over 2}-1\right]_1\cdots\,\left[{D\over
2}-1\right]_s}}=0\,. \label{comp2}
\end{eqnarray}
They are invariant under the gauge transformations
(\ref{gaugetransf}) with traceless multi-index gauge
functions  $\xi(x)$ and also require the higher spin gauge field to be
double traceless.

\subsubsection{Non--local form of the higher spin equations}
We shall now demonstrate how the higher spin field equations with
the compensator (\ref{comp}) are related to the non--local
equations of Francia and Sagnotti
\cite{Francia:2002aa,Francia:2002pt}. We shall consider the simple
(standard) example of a gauge field of spin 3. The case of a
generic spin $s$ can be treated in a similar but more tedious way.
In a somewhat different way the relation of the compensator
equations to non--local higher spin equations was discussed in
\cite{Francia:2002pt}.

For the {\it spin} 3 field the compensator equation takes the form
\begin{equation}\label{3}
G_{mnp}:=\square\,\varphi_{mnp}-3\partial_q\,\partial_{(m}\,\varphi^{~~~q}_{np)}
+3\partial_{(m}\,\partial_n\,\varphi_{p)q}^{~~~q}=\partial_m\,\partial_n\,\partial_p\,\rho(x)\,,
\end{equation}
where $()$   stands for the symmetrization of the indices with
weight one and $\rho(x)$ is the compensator, which is a scalar field
in the case of spin 3.

We now take the derivative and then the double trace of the left
and the right hand side of this equation and get
\begin{equation}\label{dg}
\partial_m\,G^{mn}_{~~~n}=\square^2\,\rho(x)\,.
\end{equation}
Modulo the {\it doubly harmonic} zero modes $\rho_0(x)$,
satisfying $\square^2\,\rho_0(x)=0$, one can solve eq. (\ref{dg})
for $\rho(x)$ in a non--local form
\begin{equation}\label{rho}
\rho(x)={1\over \square^2}\,\partial_m\,G^{mn}_{~~~n}\,.
\end{equation}
Substituting this solution into the spin 3 field equation
(\ref{3}) we get one of the forms of non-local equations
constructed in \cite{Francia:2002aa, Francia:2002pt}
\begin{equation}\label{FS}
G_{mnp}:=\square\,\varphi_{mnp}-3\partial_q\,
\partial_{(m}\,\varphi^{~~~q}_{np)}
+3\partial_{(m}\,\partial_n\,\varphi_{p)q}^{~~~q}={1\over
\square^2}\,\partial_m\,\partial_n\,\partial_p\,
(\partial_q\,G^{qr}_{~~~r})\,.
\end{equation}

Let us now consider the more complicated example of {\it spin} 4.
In the Fronsdal formulation, the fields of spin 4 and higher
feature one more restriction: they are double traceless. We shall
show how this constraint appears upon gauge fixing the compensator
equation, which for the spin 4 field has the form
\begin{equation}\label{4}
G_{mnpq}:=\square\,\varphi_{mnpq}-4\partial_r\,\partial_{(m}\,\varphi^{~~~~r}_{npq)}
+6\partial_{(m}\,\partial_n\,\varphi_{pq)r}^{~~~~r}
=4\partial_{(m}\,\partial_n\,\partial_p\,\rho_{q)}(x)\,.
\end{equation}
Taking the double trace of (\ref{4}) we have
\begin{equation}\label{dtr}
G^{mn}_{~~~mn}=3\square\,\varphi^{mn}_{~~~mn}=4\square\,\partial_m\,\rho^m\,.
\end{equation}
Taking the divergence and the trace of (\ref{4}) we get
\begin{equation}\label{ddtr}
\partial_m\,G^{mn}_{~~~np}=\square^2\rho_p+3\partial_p\,\square\,\partial_m\,\rho^m=
\square^2\rho_p+{3\over 4}\,\partial_p\,G^{mn}_{~~~mn}\,,
\end{equation}
where we have used (\ref{dtr}) to arrive at the right hand side of
(\ref{ddtr}).

  From (\ref{ddtr}) we find that modulo the zero modes $\rho^p_0$ of
$\square^2\,\rho^p_0=0$, the compensator field is non--locally
expressed in terms of the (double) trace of the Fronsdal kinetic
term
\begin{equation}\label{rhom}
\rho_p={1\over \square^2}\,(\partial_m\,G^{mn}_{~~~np}-{3\over
4}\,\partial_p\,G^{mn}_{~~~mn})\,.
\end{equation}
Inserting (\ref{rhom}) into (\ref{4}) we  get one of the forms of
 the non--local
Francia--Sagnotti equations for the spin 4 field.

Consider now the following identity
\begin{equation}\label{bi}
\partial_q\,G^{q}_{~~mnp}-\,\partial_{(m}\,G^q_{~~np)q}=-{3\over
2}\partial_m\,\partial_n\,\partial_p\,\varphi^{qr}_{~~~qr}=-{
2}\,\partial_m\,\partial_n\,\partial_p\,(\partial_q\,\rho^q)\,.
\end{equation}
  From (\ref{dtr}) and (\ref{bi}) it follows that modulo constant,
linear and quadratic terms in
$x^m$ (which can be put to zero by requiring an appropriate
asymptotic (fall--off) behavior of the wave functions at infinity)
 the double trace of the gauge field $\varphi(x)$ is
proportional to the divergence of $\rho_q(x)$
\begin{equation}\label{phirho}
\varphi^{mn}_{~~~mn}={4\over 3}\,\partial_m\,\rho^m\,.
\end{equation}
Therefore, when we partially fix the gauge symmetry by putting
$\rho_q(x)=0$, the double trace of the gauge field also vanishes
and we recover the Fronsdal formulation with the traceless gauge
parameter and the double traceless gauge field.

\subsubsection{Half integer spin fields}
 Let us generalize the previous consideration to the case of
fermions. The fermionic spin--$s$ field strength ${\cal R}^\alpha$
is the spinor--tensor
\begin{equation}\label{calRcurv}
{\cal R}^\alpha_{~_{\left[{D\over 2}\right]_1\cdots\left[{D\over
2}\right]_{s-{1\over 2}}}}(x)
\end{equation}
whose tensorial part is described by the rectangular Young tableau
${D\over 2}\times (s-{1\over 2})$. It satisfies Bianchi identities
analogous to (\ref{multi}) and can be expressed, similarly to
(\ref{fielstr}), in terms of a multi--index fermionic field
potential
\begin{equation}\label{fermi}
{\cal R}^\alpha_{~_{\left[{D\over 2}\right]_1\cdots\left[{D\over
2}\right]_{s-{1\over 2}}}}=\partial_1 \cdots \partial_{s-{1\over
2}} \, \psi^\alpha_{_{\left[{D\over
2}-1\right]_1\cdots\left[{D\over 2}-1\right]_{s-{1\over 2}}}}\,,
\end{equation}
where the fermionic conformal gauge field $\psi^\alpha(x)$  is the
spinor--tensor characterized by the rectangular Young diagram
$({D\over 2}-1)\times(s-1/2)$. The gauge transformations of
$\psi^\alpha(x)$ are similar to (\ref{gaugetransf}) with the only
difference that the gauge parameter $\xi^\alpha(x)$ is now a
spinor--tensor characterized by the diagram
$(s-1/2,\ldots,s-1/2,s-3/2)$ with $({D\over 2}-1)$ rows. The
fermionic generalization of the Damour--Deser identity is
\begin{equation}
(\g^m\,{\cal R})^\alpha{_{m\,\left[{D\over
2}-1\right]_1,\,\left[{D\over 2}\right]_2\,\cdots\left[{D\over
2}\right]_{s-{1\over 2}}}} =\partial_1\cdots \partial_{s-{3\over
2}}\, {G^\alpha}_{_{\left[{D\over 2}-1\right]_1\cdots\left[{D\over
2}-1\right]_{s-{3\over 2}}\,\left[{D\over 2}-1\right]_{s-{1\over
2}}}}\,,\label{DDgenf}
\end{equation}
where the fermionic kinetic operator $G^\alpha$ acting on the
gauge field $\psi^\alpha$ is \cite{Labastida:1986ft}
\begin{eqnarray}
{G^\alpha}_{_{\left[{D\over 2}-1\right]_1\cdots\,\left[{D\over
2}-1\right]_{s-{1\over 2}}}} =\,{\not {\! \partial}
}\psi^\alpha_{_{\left[{D\over 2}-1\right]_1\cdots\left[{D\over
2}-1\right]_{s-{1\over 2}}}} -\sum_{i=1}^{s-{1\over 2}
}\partial_i\,(\gamma^n\psi)^\alpha_{_{\left[{D\over
2}-1\right]_1\cdots\,n\left[{D\over
2}-2\right]_i\,,\cdots\,\left[{D\over 2}-1\right]_{s-{1\over
2}}}}\,.
\end{eqnarray}

The field strength (\ref{fielstr}) is invariant under the
following gauge transformations of the gauge potential
\cite{DuboisV}
\begin{eqnarray}
\delta \psi^\alpha_{~_{\left[{D\over
2}-1\right]_1\cdots\,\left[{{D\over 2}}-1\right]_{s-{1\over 2}}}}
 = \sum_{i=1}^{s}\,\partial_{i}\, \xi^\a_{_{\left[{D\over 2}-1\right]_1\,\cdots
 \,\left[{D\over 2}-2\right]_i\cdots\,\left[{D\over 2}-1\right]_{s-{1\over 2}}}}\,. \label{fgaugetransf}
\end{eqnarray}

When the fermionic field strength satisfies the
$\g$--tracelessness condition (iv) of Sec. \ref{conformalHS}, {\it
i.e.}
\begin{equation}\label{gtrace}
(\g^m\,{\cal R})^\alpha{_{m\,\left[{D\over
2}-1\right]_1,\,\left[{D\over 2}\right]_2\,\cdots\left[{D\over
2}\right]_{s-{1\over 2}}}} =0\,,
\end{equation}
eq. (\ref{DDgenf}) implies that $G^\alpha$ is $\partial^{s-{3\over
2}}$--closed. Since $\partial^{s+{1\over 2}}\equiv 0$, by virtue of
the generalized Poincar\'e lemma $G^\alpha$ is $\partial^2$--exact
\begin{equation}
{G^\alpha}_{_{\left[{D\over 2}-1\right]_1\cdots\,\left[{D\over
2}-1\right]_{s-{1\over 2} }}}=\sum_{j>i=1}^{s-{1\over 2}
}\,\partial_i\,\partial_j\, \rho^\alpha_{~_{\left[{D\over
2}-1\right]_1\cdots\,,\left[{D\over
2}-2\right]_{i}\,,\cdots\,,\left[{D\over
2}-2\right]_{j}\,,\cdots\,\left[{D\over 2}-1\right]_{s-{1\over 2
}}}}\,, \label{compf}
\end{equation}
where $\rho^\alpha(x)$ is the fermionic compensator characterized
by a Young diagram $(s-1/2,\ldots,s-1/2,s-5/2)$ with $({D\over
2}-1)$ rows.

Equation (\ref{compf}) is the generalization of the compensator
equation given in \cite{Sagnotti:2003qa} to arbitrary rectangular
Young diagrams. The demonstration of its relation to the
gamma--traceless part of the fermionic higher spin field strength
is a new result.

The gauge variation of $G^\alpha(x)$ is
\begin{equation}\label{GGg}
\d{G^\alpha}_{_{\left[{D\over 2}-1\right]_1\cdots\,\left[{D\over
2}-1\right]_{s-{1\over 2}}}} = \sum_{j>i=1}^{s-{1\over 2}
}\,\partial_i\,\partial_j\,
(\gamma^{m}\,\xi)^\alpha_{~_{\left[{D\over
2}-1\right]_1\cdots\,,\left[{D\over
2}-2\right]_{i}\,,\cdots\,,\left[{D\over 2}-2\right]_{j}\,
m,\cdots\,,\cdots\,\left[{D\over 2}-1\right]_{s-{1\over 2}}}}\,
\end{equation}
and it is compensated by a gauge shift of the field
$\rho^\alpha(x)$ given by the $\gamma$--trace ({\it i.e.} the
contraction of the gamma matrix vector index with one inside a
cumulative index) of the gauge parameter
\begin{equation}\label{rhogf}
\delta\rho^\alpha_{~_{\left[{D\over 2}-2\right]_{1}\,\left[{D\over
2}-2\right]_{2}\, \left[{D\over
2}-1\right]_3\,\cdots\,\left[{D\over 2}-1\right]_{s-{1\over 2}
}}}=(\gamma^{m}\,\xi)^\alpha_{_{\left[{D\over
2}-2\right]_{1}\,m\,,\,\left[{D\over 2}-2\right]_{3}\,
\left[{D\over 2}-1\right]_3\,\cdots\,\left[{D\over
2}-1\right]_s}}\,.
\end{equation}
Thus, the compensator can be gauged away by choosing a gauge
parameter $\xi^\alpha(x)$ with the appropriate $\gamma$--trace.
Then, the equations of motion of the gauge field $\psi^\alpha(x)$
become the first order differential equations of Labastida which
generalize to mixed symmetry fields those of Fang and Fronsdal
\begin{equation}\label{lf}
\,{\not {\! \partial} }\psi^\alpha_{_{\left[{D\over
2}-1\right]_1\cdots\left[{D\over 2}-1\right]_{s-{1\over 2}}}}
-\sum_{i=1}^s\partial_i\,(\gamma^n\psi)^\alpha_{_{\left[{D\over
2}-1\right]_1\cdots\,,\,n\left[{D\over
2}-2\right]_i\,,\cdots\,\left[{D\over 2}-1\right]_{s}}}=0\,.
\end{equation}
These equations are invariant under the gauge transformations
(\ref{fgaugetransf}) with $\gamma$--traceless parameters.

Alternatively, one can get the non--local Francia--Sagnotti
equations for fermions by taking a particular non--local solution
for the compensator field in terms of the  fermionic kinetic
operator $G^\alpha$. As a simple example consider the $s=5/2$
case. Eq. (\ref{compf}) takes the form
\begin{equation}\label{5/2}
G^\alpha_{mn}:={\not {\! \partial}
}\psi^\alpha_{mn}-2\partial_{(m}\,(\gamma^q\psi)^\alpha_{n)q}
=\partial_m\,\partial_n\,\rho^\alpha(x)\,.
\end{equation}
Taking the trace of (\ref{5/2}) we get
\begin{equation}\label{frho}
\square\,\rho^\alpha=G^{\alpha p}_{~~~p}\,.
\end{equation}
Hence, modulo the zero modes $\rho^\alpha_0(x)$ of the
Klein--Gordon operator $\square\,\rho^\alpha_0=0$ the compensator
field is non--locally expressed in terms of the trace of
$G^\alpha_{mn}$
\begin{equation}\label{nfrho}
\rho^\alpha={1\over \square}\,G^{\alpha p}_{~~~p}\,.
\end{equation}
Substituting (\ref{nfrho}) into (\ref{5/2}) we get the
Francia--Sagnotti equation for the fermionic field of spin 5/2
\begin{equation}\label{FS5/2}
G^\alpha_{mn}:={\not {\! \partial}
}\psi^\alpha_{mn}-2\partial_{(m}\,(\gamma^q\psi)^\alpha_{n)q}
={1\over \square}\,\partial_m\,\partial_n\,G^{\alpha p}_{~~~p}\,.
\end{equation}

In the same way one can relate the compensator equations for an
arbitrary half integer spin field to the corresponding non--local
field equation. As in the bosonic case, one can find that for
$s\geq {7\over 2}$ the triple--gamma trace of the fermionic gauge
field potential is expressed in terms of the $\gamma$--trace and
the divergence of the compensator field.

\setcounter{equation}0\section{Dynamics of
the tensorial twistor--like particle. Preonic equation
and conformally invariant fields}\label{Preonicequ}

We now show that the conformal integer and half integer higher
spin fields in 4--, 6-- and 10--dimensional space--time satisfying
the geometrical equations considered in the previous Section arise
  as a result of the quantization of a twistor--like
particle propagating,
 respectively, in the $n=4$, 8 and 16 tensorial spaces. The
quantum spectrum of this particle contains an infinite number of
conformal higher spin states. The state of each spin appears only
once in the spectrum of $n=4$ ($D=4$) and $n=16$ ($D=10$)
tensorial particles, while (except for the scalar and the spinor
state) the higher spin fields in the spectrum of the $n=8$
tensorial particle form in the corresponding $D=6$ space--time
higher {\it isospin} representations of an internal group $SO(3)$.
Let us however note that the considerations below (up to eq.
(\ref{f})) are valid for arbitrary $n$.

 The action proposed in \cite{Bandos:1998vz} to describe a
twistor--like particle propagating in tensorial space has the form
\begin{equation}\label{action}
S[X,\l]=\int\,
E^{\a\b}\left(X(\tau)\right)\,\l_\a(\tau)\,\l_\b(\tau),
\end{equation}
where $\l_\a(\tau)$ is an auxiliary commuting real spinor, a
\textit{twistor--like} variable, and $E^{\a\b}(x(\tau))$ is the pull
back on the particle worldline of the tensorial space vielbein. In
this paper we will deal with flat tensorial space. In this case
\begin{equation}\label{Omega}
E^{\a\b}(X(\tau))=d\tau\,\partial_\tau
X^{\a\b}\,(\tau)=dX^{\a\b}\,(\tau)\,.
\end{equation}
The dynamics of particles on the supergroup manifolds
$OSp(N|n,\mathbb R)$ (which are the tensorial extensions of AdS
superspaces) was considered for $N=1$ in
\cite{blps,Plyushchay:2003gv,Plyushchay:2003tj} and for a generic
$N$ in \cite{Vasiliev:2001zy,Misha}. The twistor--like
superparticle in $n=32$ tensorial superspace was considered in
\cite{Bandos:2003us} as a point--like model for BPS preons
\cite{Bandos:2001pu}, the hypothetical
${{31}\over{32}}$--supersymmetric constituents of M--theory.

The action (\ref{action}) is manifestly invariant under global
$GL(n,\mathbb R)$ transformations. Without going into details
which the reader may find in
\cite{Bandos:1998vz,Vasiliev:2001zy,Plyushchay:2003gv}, let us
note that the action (\ref{action}) is invariant under global
$Sp(2n,\mathbb R)$ transformations, acting non--linearly on
$X^{\a\b}$ and on $\l_\a$, {\it i.e.} it possesses the symmetry
considered by Fronsdal to be an underlying symmetry of higher spin
field theory in the case $n=4$, $D=4$ \cite{fronsdal1}.

Applying the Hamiltonian analysis to the particle model described by
(\ref{action}) and (\ref{Omega}), one finds that the momentum
conjugate to $X^{\a\b}$ is related to the twistor--like variable
$\l_\a$ via the constraint
\begin{equation}\label{Penroselike}
P_{\alpha\beta}=\lambda_\alpha\lambda_\beta\,.
\end{equation}
This expression is the direct analog and generalization of the
Cartan--Penrose (twistor) relation for the particle momentum
$P_m=\lambda\gamma_m\lambda$. In virtue of the Fierz identity
(\ref{D101}) the twistor particle momentum is light--like in
$D=3,4,6$ and $10$ space--time. Therefore, in the tensorial spaces
corresponding to these dimensions of space--time the
first--quantized particles are massless
\cite{Bandos:1998vz,Bandos:1999qf}.

The quantum counterpart of (\ref{Penroselike}) is the equation
\cite{Bandos:1999qf}
\begin{equation}\label{l}
D_{\a\b}\Phi(X,\l)=\left({\partial\over{\partial
X^{\a\b}}}-i\l_\a\l_\b\right)\Phi(X,\l)=0\,,
\end{equation}
where the wave function $\Phi(X,\l)$ depends on $X^{\a\b}$ and
$\l_\a$.   Eq. (\ref{l}) has been shown to correspond
\cite{Bandos:2003us} to a BPS preon \cite{Bandos:2001pu} and thus
may be called {\it preonic equation}. The general solution of
(\ref{l}) is the plane wave
\begin{equation}\label{soll}
\Phi(X,\l)=e^{iX^{\a\b}\l_\a\l_\b}\varphi(\l),
\end{equation}
where $\varphi(\l)$ is a generic function of $\l_\a$.

One can now Fourier transform the function (\ref{soll}) to another
representation to be called $Y$--representation
\begin{equation}\label{yr}
C(X,Y)=\int\,d^4\l\,e^{-iY^\a\l_\a}\Phi(X,\l)=\int\,d^4\l\,e^{-iY^\a\l_\a+
i X^{\a\b}\l_\a\l_\b}\varphi(\l).
\end{equation}
The wave function $C(X,Y)$ satisfies the Fourier transformed
preonic equation
\begin{equation}\label{Y}
\left({\partial\over{\partial
X^{\a\b}}}+i{\partial^2\over{\partial Y^\a\partial
Y^\b}}\right)C(x,Y)=0.
\end{equation}
This equation has been analyzed in detail in \cite{Vasiliev:2001zy}
for wave functions that are power series in $Y^\a$
\begin{equation}\label{pol}
C(X,Y)=\sum^\infty_{n=0}C_{\a_1\cdots\a_n}(X)\,Y^{\a_1}\cdots
Y^{\a_n}=b(X)+f_\a(X)Y^\a+\cdots\,.
\end{equation}
In view of the Fourier relation (\ref{yr}) the series in $Y^\a$
naturally arises as a result of the series expansion of the
exponent $e^{-iY^\a\l_\a}$. Thus the scalar field $b(X)$ and the
spinor field $f_\a(X)$ in ({\ref{pol}) are related to the wave
function $\Phi(x,\l)$ by the following integral expressions
\begin{eqnarray}b(X)&=&\int d^{n}\l\,\,\Phi(X,\l)\,+
{~~~c.c.}\,,\label{bis}\\
f_\a(X)&=&-i\int d^{n}\l\,\,\l_\a\,\Phi(X,\l)\,+
{~~~c.c.}\,,\label{fis}
\end{eqnarray}
where $c.c.$ stands for `complex conjugate' since in what follows
we shall deal with the real fields $b(X)$ and $f_\a(X)$.

Inserting (\ref{pol}) into (\ref{Y}) one finds that the scalar
field $b(X)$ and the spinor field $f_\a(X)$ must satisfy the
following equations found in \cite{Vasiliev:2001zy}
\begin{eqnarray}\label{b}
\partial_{\a\b}
\partial_{\g\d}\,b(X)-\partial_{\a\g}\partial_{\b\d}\,b(X)&=&0\,,\\
\quad \partial_{\a\b} f_\g(X)-\partial_{\a\g}
f_\b(X)&=&0\label{f}\,.
\end{eqnarray}
These fields are dynamical, while all higher components in the
expansion (\ref{pol}) are expressed in terms of (higher)
derivatives of the basic fields $b(X)$ and $f_\a(X)$ and, hence,
are auxiliary fields \cite{Vasiliev:2001zy}. In
\cite{Vasiliev:2001zy} it was also shown that eqs. (\ref{b}) and
(\ref{f}) are invariant under the generalized superconformal
transformations generating the supergroup $OSp(1|2n)$. The fields
$b(X)$ and $f_\a(X)$ form a linear supermultiplet (a
supersingleton) of a subgroup of $OSp(1|2n)$ acting linearly in
the tensorial superspace. The superfield form of the equations
(\ref{b}) and (\ref{f}), both on flat tensorial superspace and on
the supergroup manifold $OSp(1|n)$, have been constructed in
\cite{Bandos:2004nn}.

The general solutions of the equations (\ref{b}) and (\ref{f}) are
eqs. (\ref{bis}) and (\ref{fis}) with $\Phi(X,\lambda)$ being the
plane wave (\ref{soll}). They will prove to be useful for the
derivation  of the geometrical higher spin equations in $D=4,6$ and
$10$ space--time from the tensorial equations (\ref{b}) and
(\ref{f}).

Let us recall that $X^{\a\b}$ stands for the tensorial coordinates
containing the conventional space--time coordinates $x^m$ and the
`helicity' degrees of freedom $y^{m\cdots p}$. In order to make
 contact with the ordinary space--time picture, one has to
single out the $y$-dependence of $b(X)$ and $f_\a(X)$ using the
decomposition (\ref{xab}). For instance, using the form of the
general solution (\ref{soll}),
\begin{equation}\label{sollD}
\Phi(X,\l)=\Phi(x,y,\l)\,e^{ix^m\,\l \gamma^m\l}\,e^{iy^{m\cdots p
}\,\l\g_{m\cdots p}\l}\,\varphi(\l)
\end{equation}
(where the contraction of the spinor indices is implied, {\it
e.g.} $ \l\g^m\l\equiv \l^\a \g^m_{\a\b}\l^\b$), one finds that in
view of (\ref{bis}), (\ref{fis}) and the Fierz identity
\begin{equation}\label{usef103}
(\gamma^m\l)_\a(\l\g_m\l)=0\,
\end{equation}
for $D=3,4,6$ and 10, the fields $b(x,y)$ and $f_\alpha(x,y)$
satisfy the massless Klein--Gordon equation
\begin{equation}\partial^m\partial_m
b(x^p,y)=0\,,\label{KGD}\end{equation}
 and the Dirac equation
\begin{equation}
\g^m\partial_m f(x^p,y)=0\,.\label{WeylD}
\end{equation}

We shall now turn to a more detailed analysis of the tensorial
equations (\ref{b}) and (\ref{f}) and of their relation to the
geometrical equations for the conformal higher spin fields in
$D=3,4,6$ and 10 space--time.

\setcounter{equation}0\section{How the
quantum dynamics of the
tensorial particle produces conformal higher spin
fields}\label{Spectrum}

\subsection{n=2, D=3}

This case is very simple because there are no  extra `helicity'
coordinates: the $y$ variable is absent. This is because a
complete basis of symmetric $2\times 2$ matrices is formed by the
symmetric $D=3$ Dirac matrices $\g^m_{\a\b}$. Hence,
$$X^{\a\b}=\g_m^{\a\b} x^m\,\quad \Leftrightarrow \quad
x^m=\frac12 \gamma^m_{\a\b}\,X^{\a\b} \,, \quad(\a,\b=1,2\,;\quad
m=0,1,2)\,,$$ and $b(X^{\a\b})$ and $f_\a(X^{\a\b})$ are simply the
$D=3$ space--time scalar $b(x^m)$ and spinor $f_\a(x^m)$ fields.

The only antisymmetric matrix is $\varepsilon_{\alpha\beta}$ playing
the role of the charge conjugation matrix. It can be used to lower
and raise the spinor indices. Therefore, (\ref{f}) is equivalent to
the Dirac equation
$\partial_{\a\b}f^\b(x^m)=(\gamma^m\,\partial_m\,f)_\alpha=0$,
while (\ref{b}) is equivalent to the massless Klein-Gordon equation
$\square\, b(x^m)=0$. These two equations provide the complete set
of $D=3$ Poincar\'e group unitary irreps extendable to unitary
representations of the $D=3$ conformal group which, via the
isomorphism
$Spin(2,3)\cong Sp(4,\mathbb R)$, coincides with the symmetry
group $Sp(4,\mathbb R)$ of the $n=2$ tensorial space.

\subsection{n=4, D=4}

\subsubsection{Coordinates}

The ten-dimensional tensorial space is parametrized by
\begin{equation}\label{x}
X^{\a\b}={1\over 2}\,x^m\gamma_m^{\a\b}+{1\over 4}\,
y^{mn}\gamma_{mn}^{\a\b}\,, \quad
(m,n=0,1,2,3\,;\,\,\a,\b=1,2,3,4)\,,
\end{equation}
where $x^m= 1/2\,X^{\a\b}\gamma^m_{\a\b}$ are associated with the
four coordinates of  conventional $D=4$ space--time and the six
$y^{mn}=1/2\,X^{\a\b}\gamma^{mn}_{\a\b}$ that describe the spin
degrees of freedom. The derivative with respect to $X^{\a\b}$ is
\begin{equation}
\partial_{\a\b}={1\over 2}\,\gamma^m_{\a\b}\,\partial_m+{1\over
2}\,
\gamma^{mn}_{\a\b}\,\partial_{mn}\,,\label{derx}\end{equation}
where $\partial_m$ and $\partial_{mn}$ are the derivatives along
$x^m$ and $y^{mn}$, respectively.

\subsubsection{Unfolded equations}\label{unfoldsect}

Let us now make a short digression and recall the unfolded
formulation of $D=4$ higher-spin fields, which is usually
constructed using the two component Weyl spinors (see
\cite{Vasiliev:2001zy,V01} for details and references). The
Majorana spinor index $\a$ is then decomposed into a pair of Weyl
indices $\a=(A,\dot{A})$ with $A,\dot{A}=1,2$ and
$\l^\a=(\l^A\,,\overline{\l}^{\dot A} )$ with
$(\l^A)^*=\overline{\l}^{\dot A}\,.$ The momentum constraint
(\ref{Penroselike}) takes the form
\begin{eqnarray}
P_{AB}=\l_A\l_B\,,\quad \overline{P}_{\dot{A}\dot{B}}=
\overline{\l}_{\dot A}\overline{\l}_{\dot B}\,,
\quad P_{A\dot{A}}=\l^{}_A\overline{\l}_{\dot A}\,,
\end{eqnarray}
where the last equation  is the Cartan--Penrose representation of
the light-like momentum. In the same manner, the preonic equation
in the $Y$--representation, eq. (\ref{Y}), splits into
\begin{eqnarray}\label{Yy}
\left(\sigma^{mn}_{AB}{\partial\over{\partial
y^{mn}}}+i{\partial^2\over{\partial Y^A\partial
Y^B}}\right)C(x,y,Y)=0,\,\nonumber\\
\\
\left(\overline{\sigma}^{mn}_{{\dot A}{\dot
B}}{\partial\over{\partial y^{mn}}}-i{\partial^2\over{\partial
\overline{Y}^{\dot A}\partial \overline{Y}^{\dot
B}}}\right)C(x,y,Y)=0\nonumber
\end{eqnarray}
and
\begin{eqnarray}\left(\sigma^m_{A\dot{A}}{\partial\over{\partial
x^m}}+i{\partial^2\over{\partial Y^A\partial \bar{Y}^{\dot
A}}}\right)C(x,y,Y)=0\,,\label{unfold}
\end{eqnarray}
where $\sigma^m_{A\dot{A}}$ are the Pauli matrices and
$\sigma^{mn}_{AB}=\sigma^{[m}_{A\dot{A}}\,\sigma^{n]\dot A}_{B}$.

Equations (\ref{Yy}) relate the dependence of $C(x,y,Y)$ on
$y^{mn}$ to its dependence on $Y^\a$. Using this relation one can
regard the wave function
$C(x^m,Y^\a):=C(X^{\a\b},Y^\a)|_{y^{mn}=0}$ at $y^{mn}=0$ as the
fundamental field and thus arrives at the unfolded formulation of
\cite{V01} whose basic equation for $C(x^m,Y^\a)$ is (\ref{unfold}).

The consistency of (\ref{unfold}) implies the integrability
conditions
\begin{eqnarray}
\label{masseq}\frac{\partial^2}{\partial Y^{[A}
\partial
x^{B]\dot B}}\, C(x^{C{\dot C}},Y) =0\,,\,
\frac{\partial^2}{\partial \bar{Y}^{[\dot A} \partial x^{{\dot
B}]B}}\, C(x^{C\dot C},Y) =0\,.
\end{eqnarray}

The expansion of $C(x^m,Y)$ in terms of $Y^A$ and
$\overline{Y}^{\dot A}$
is\begin{equation}C(x^p,Y^A,\overline{Y}^{\dot
A})=\sum_{m,n=0}^{\infty}\frac{1}{m!n!}\, C_{A_1 \ldots
A_m,\,{\dot B}_1 \ldots {\dot B}_n }(x^p)\, Y^{A_1} \ldots Y^{A_m}
\,\overline{Y}^{{\dot B}_1} \ldots \overline{Y}^{{\dot B}_n}\,,
\end{equation}
where  reality  imposes $(C_{A_1 \ldots A_m,\,{\dot B}_1 \ldots
{\dot B}_n })^*=C_{B_1 \ldots B_n,\,{\dot A}_1 \ldots {\dot A}_m }$,
and the spin--tensors $C$ are by definition symmetric in the indices
$A_i$ and in
$\dot B_i$. All the components of $C(x^m,Y^A,\overline{Y}^{\dot
A})$ that depend on {\it both} $Y^A$ and $\overline{Y}^{\dot A}$
are auxiliary fields expressed by (\ref{unfold}) in terms of
space--time derivatives of the dynamical fields contained in the
analytic fields $C(x^m,Y^A,0)$ and $C(x^m,0,Y^{\dot A})$. The only
dynamical fields are the self--dual and anti--self--dual
components $C_{A_1\ldots A_{2s}}(x^m)$ and $C_{{\dot A}_1\ldots
{\dot A}_{2s}}(x^m)$ of the spin--$s$ field strength. The
nontrivial equations on the dynamical fields are \cite{V01} the
Klein--Gordon equation for the spin zero scalar field  $\square
C=0$  and the massless Bargmann--Wigner equations \cite{Bargmann}
for spin $s>0$ field strengths
\begin{equation}\partial^{B\dot B}C_{BA_1\ldots A_{2s-1}}(x)=0\,,\quad
\partial^{B\dot B}C_{{\dot B}{\dot A}_1\ldots {\dot A}_{2s-1}}(x)=0\,,\label{BW}
\end{equation}
which follow from (\ref{masseq}) \footnote{The well known counting
of the degrees of freedom is as follows: the symmetric tensor
$C_{BA_1\ldots A_{2s-1}}$ has ${2s+1\choose 2s}=2s+1$ components
satisfying  ${{2s}\choose{2s-1}}=2s$ independent conditions; this
leaves in $C_{BA_1\ldots A_{2s-1}}$ one independent helicity
degree of freedom, as is well known for the massless spin $s$
fields in $D=4$. The spin $s$ state of opposite helicity is
described by $C_{{\dot B\dot A}_1\ldots {\dot A}_{2s-1}}(x^m)$.}.

The massless $D=4$ higher spin field equations are known to be
conformally invariant. So the $SU(2,2)$ symmetry of this infinite
set of massless relativistic equations in $D=4$ gets extended to the
$OSp(1|8,\mathbb R)$ symmetry that becomes more transparent in the
tensorial space \cite{Vasiliev:2001zy}.

Note that $C_{A_1\ldots A_{2s}}(x^m)$ and $C_{{\dot A}_1\ldots
{\dot A}_{2s}}(x^m)$ are related to the integer spin curvature
tensor (\ref{Rcurv}) and to the half integer spin curvature
(\ref{calRcurv}) in $D=4$ as follows
\begin{equation}\label{RC}
R_{m_1n_1,\cdots,\,
m_sn_s}=\sigma_{m_1\,n_1}^{A_1A_{s+1}}\cdots\,\sigma_{m_s\,n_s}^{A_sA_{2s}}
\,C_{A_1\cdots\,A_s\,A_{s+1}\cdots\, A_{2s}}+ ~c.c.
\end{equation}
\begin{equation}\label{calRC}
{\cal R}^{A_{2s}}_{m_1n_1,\cdots,\,
m_{s-{1\over 2}}n_{s-{1\over 2}}}=\sigma_{m_1\,n_1}^{A_1A_{s+{1\over 2}}}
\cdots\,\sigma_{m_{s-{1\over 2}}\,n_{s-{1\over 2}}}^{A_{s- {1\over 2}}A_{2s-1}}
\,C_{A_1\cdots\,A_{s-{1\over 2}}\,A_{s+{1\over 2}}\cdots\,
A_{2s-1}}^{A_{2s}}
\end{equation}

\subsubsection{The geometric equations from the scalar
and spinor field equations in tensorial space}

Alternatively to the unfolded construction of Sec. 4.2.2 where we
kept the dependence of the wave function on $x^m$ and $Y^\a$ and
effectively eliminated its dependence on $y^{mn}$, one can deal with
the fields $b(x^l,\,y^{mn})$ and $f_\a(x^l,\,y^{mn})$ and their
field equations (\ref{b}) and (\ref{f}) in tensorial space
\cite{Vasiliev:2001zy}. This formulation will prove  to be more
convenient for higher--dimensional generalizations.

Since in $D=4$ the set
$\{C^{\b\g},\gamma_5^{\b\g},(\g_5\g_p)^{\b\g}\}$ (where $C^{\b\g}$
is the charge conjugation matrix and
$\gamma_5=\gamma_0\gamma_1\gamma_2\gamma_3$, $(\gamma_5)^2=-1$)
forms   a basis of $4\times 4$ antisymmetric matrices, the equation
of motion (\ref{f}) of the tensorial space field
$f_\a(X)$ is equivalent to the system of {\it linearly dependent}
differential equations
\begin{eqnarray}\label{f0}
(\g^m\partial_m-\g^{mn}\partial_{mn})f=0,\nonumber\\
(\g^m\partial_m-\g^{mn}\partial_{mn})\g_5f=0,\\
(\g^m\partial_m-\g^{mn}\partial_{mn})\g_5\g_p\,f=0\,,\nonumber
\end{eqnarray}
where the expression (\ref{derx}) for the tensorial partial
derivative has been used.

 Moving $\gamma_5$ and $\g_5\g_p$ to the left hand side of (\ref{f0})
and taking linear combinations of the resulting equations, one
gets the following equivalent set of {\it independent} equations
\begin{eqnarray}
\gamma^p\,\partial_{p}\,f(x^l,y^{mn})=0\,, \label{Dirac}\\
\nonumber\\
\left({\partial_p}-2\gamma^r\,{\partial_{rp}}\right)f(x^l,y^{mn})=0\,.
\label{gammatrace}
\end{eqnarray}

  {}From eqs. (\ref{Dirac}) and (\ref{gammatrace}) one can
derive the equation
\begin{equation}\label{split}
\partial_{mn}\,f={1\over 2}\,(\partial_{mn}+{1\over
2}\epsilon_{mnpq}\,\partial^{pq}\,\gamma_5)\,f+{1\over
2}\,\gamma_{[m}\,\partial_{n]}\,f\,,
\end{equation}
which describes the decomposition of the spinor-tensor
$\partial_{mn}\,f$ into the self--dual gamma--traceless part $(\partial_{mn}+{1\over
2}\epsilon_{mnpq}\,\partial^{pq}\,\gamma_5)\,f$
\begin{eqnarray}\label{trlsd}
\gamma^m\,(\partial_{mn}+{1\over
2}\epsilon_{mnpq}\,\partial^{pq}\,\gamma_5)\,f=0, \nonumber\\
(\partial_{mn}+{1\over
2}\epsilon_{mnpq}\,\partial^{pq}\,\gamma_5)\,f={1\over
2}\,\epsilon_{mnrs}\,\gamma_5\,(\partial^{rs}+{1\over
2}\epsilon^{rspq}\,\partial_{pq}\,\gamma_5)\,f
\end{eqnarray}
and the `tracefull' part  which is proportional to the $D=4$
space--time derivative of $f(x,y)$, {\it i.e.} ${1\over
2}\,\gamma_{[m}\,\partial_{n]}\,f$.

 Analogously to the fermionic equations, the equation of motion
(\ref{b}) of the tensorial space scalar $b(x,y)$ is equivalent to
\begin{eqnarray}
\partial_p\,\partial^p\,b(x^l,y^{mn})=0\label{KG}\,,\\
\nonumber\\
\left(\partial_p\,\partial_q
-4\,\partial_{pr}\,\partial^r_{~q}\right)\,b(x^l,y^{mn})=0\,,\label{trace}\\
\nonumber\\
\epsilon^{pqrt}\partial_{pq}\,\partial_{rs}\,b(x^l,y^{mn})=0\,,\label{Bianchi1}\\
\nonumber\\
\epsilon^{pqrt}\partial_q\,\partial_{rt}\,\,b(x^l,y^{mn})=0\,,\label{Bianchi2}\\
\nonumber\\
\partial^{~p}_{q}\,\partial_p\,\,b(x^l,y^{mn})=0\,. \label{transversality}
\end{eqnarray}
Roughly speaking, the system of equations
(\ref{KG})-(\ref{transversality}), which also holds for the spinor
field, is the ``square" of eqs. (\ref{Dirac})-(\ref{gammatrace}),
because the former can be obtained from the latter as
integrability conditions and using the duality relation
\begin{equation}
\g_{mn}=\frac{1}{2}\,\epsilon_{mnpq}\,\g_5\g^{pq}\,. \label{selfd4}
\end{equation}

For further generalization to higher dimensions $D=6$ and $D=10$
it is instructive to derive equations
(\ref{Dirac})--(\ref{transversality}) by applying the derivatives
$\partial_m$ and $\partial_{mn}$ to the general solutions
(\ref{bis}) and (\ref{fis}) of the tensorial equations (\ref{b}) and
(\ref{f}) and using $\gamma$--matrix Fierz identities.


 This way of deriving  the Dirac (\ref{Dirac})
and Klein--Gordon (\ref{KG}) equations has already been explained
at the end of Sec. 3, so we proceed with the consideration of the
other equations.

To get (\ref{gammatrace}) we take the derivative
$\gamma^r\,{\partial_{rp}}$ of (\ref{fis}), where
$\Phi(X,\lambda)$ is the plane wave (\ref{sollD}), and notice that
\begin{equation}\label{fierz1}
2(\lambda\gamma^r)^\a\,(\lambda\gamma_{rp}\lambda)=\lambda^\a\,(\lambda\gamma_{p}\lambda)\,
\end{equation}
holds due to the well known  Fierz identity in $D=3,4,6$ and $10$
\begin{equation}\label{basicfier}
\gamma_{m\,(\a\b}\,\gamma^m_{\g)\d}=0\,.
\end{equation}

Eq. (\ref{trace}) is obtained by taking and comparing the second
derivatives of (\ref{bis}) and (\ref{sollD}), and noticing that,
as a consequence of eq. (\ref{fierz1}),
\begin{equation}\label{fierz2}
 4\,(\lambda\gamma^{mp}\lambda)\,
(\lambda\gamma_p^{~n}\lambda)=(\lambda\gamma^m\lambda)\,(\lambda\gamma^n\lambda)\,.
\end{equation}
In the same way one checks that eqs. (\ref{Bianchi1}),
(\ref{Bianchi2}) and (\ref{transversality}) hold, respectively,
due to the algebraic identities
\begin{eqnarray}\label{bi1}
&\epsilon_{mnpq}\,(\lambda\gamma^{mn}\lambda)\,(\lambda\gamma^{pq}\lambda)=
2\,(\lambda\gamma_5\gamma^{mn}\lambda)\,(\lambda\gamma_{mn}\lambda)=0\,,\\
\nonumber\\
\label{bi2}
&\epsilon_{mnpq}\,(\lambda\gamma^{mn}\lambda)\,(\lambda\gamma^{p}\lambda)
=2\,(\lambda\gamma_5\gamma^{mn}\lambda)\,(\lambda\gamma_{n}\lambda)=0\,,\\
\nonumber\\
\label{t2}
 &(\lambda\gamma^m\lambda)\,(\lambda\gamma_{mn}\lambda)=0\,.
\end{eqnarray}
Note that all the identities (\ref{fierz2})--(\ref{t2}) are
consequences of (\ref{fierz1}). This explains from the
twistor--like point of view why the set of  eqs. (\ref{Dirac}) and
(\ref{gammatrace}) is the ``square root" of
(\ref{KG})--(\ref{transversality}).

Let us now analyze the physical meaning of the equations
(\ref{Dirac})--(\ref{transversality}) from the point of view of
the effective four--dimensional field theory. As we shall show,
eqs. (\ref{Bianchi1}), (\ref{Bianchi2}) and (\ref{transversality})
produce, respectively, the first (eq. (\ref{i})) and the second
(eq. (\ref{closed})) Bianchi identities, and the transversality
condition (\ref{trans}) for the $D=4$ higher spin curvatures.

To this end let us expand $b(x,\,y)$ and $f_\a(x,y)$ in power
series\footnote{Such an expansion is justified by the presence in
the general solutions (\ref{bis}) and (\ref{fis}) of the tensorial
equations (\ref{b}) and (\ref{f}) of the plane wave function
(\ref{sollD}) which allows us to expand
$e^{i\l\gamma_{mn}\l\,y^{mn}}$ in power series.} of $y^{mn}$
\begin{eqnarray}
b(x,\,y)&=\phi(x)+y^{m_1n_1}F_{m_1n_1}(x)
+y^{m_1n_1}\,y^{m_2n_2}\,{\hat R}_{m_1n_1,m_2n_2}(x)\nonumber\\
&+\sum_{s=3}^{\infty}\,y^{m_1n_1}\cdots y^{m_sn_s}\,{\hat
R}_{m_1n_1,\cdots,m_sn_s}(x)\,,\label{is}
\end{eqnarray}
\begin{equation}
f^\a(x,y)=\psi^\a(x)+y^{m_1n_1}\,{\hat{\cal R}}^\a_{m_1n_1}(x )
+\sum_{s={5\over 2}}^{\infty}\,y^{m_1n_1}\cdots y^{m_{s-{1\over
2}} n_{s-{1\over 2}}}\,{\hat{\cal
R}}^\a_{m_1n_1,\cdots,m_{s-{1\over 2}}n_{s-{1\over 2}}}(x )\,.
\label{his}
\end{equation}
In the multi index notation of Sec. \ref{conformalHS} these series
take the form
\begin{equation}
b(x,\,y) =\phi(x)+y^{^{[2]}}F_{_{[2]}}(x)
+y^{^{[2]_1}}\,y^{^{[2]_2}}\,{\hat R}_{_{[2]_1 [2]_2 }}(x)
 +\sum_{s=3}^{\infty}\,y^{^{[2]_1}}\cdots
y^{^{[2]_s}}\,{\hat R}_{_{[2]_1 \cdots\,[2]_s }}(x)\,,\label{ism}
\end{equation}
\begin{equation}
 f^\a(x,y)
=\psi^\a(x)+y^{^{[2]}}\,{\hat{\cal R}}^\a_{_{[2]}}(x )
 +\sum_{s={5\over 2}}^{\infty}\,y^{^{[2]_1}}\cdots
y^{^{[2]_{s-{1\over 2}}}}\,{\hat{\cal R}}^\a_{_{{[2]_1}
\cdots\,[2]_{s-{1\over 2}}}}(x )\,. \label{hism}
\end{equation}

The scalar field $\phi(x)$ and the spinor field $\psi^\a(x)$ as
well as all the higher order tensors and spin tensors in
(\ref{is}) and (\ref{his}) satisfy the Klein--Gordon equation
(\ref{KG}) and hence are massless $D=4$ fields. The fermionic
fields $\psi^\a(x)$ and ${\hat{\cal R}}^\a_{_{{[2]_1}
\cdots\,[2]_{s-{1\over 2}}}}(x )$ satisfy the Dirac equation
(\ref{Dirac}).

Eq.  (\ref{gammatrace})  tells us that the gamma--trace of the
fermionic spin $s$ tensor ${\hat{\cal R}}^\a_{_{{[2]_1}
\cdots\,[2]_{s-{1\over 2}}}}(x )$ is proportional to the
space--time derivative of the spin $(s-1)$ tensor ${\hat{\cal
R}}^\a_{_{{[2]_1} \cdots\,[2]_{s-{3\over 2}}}}(x )$, that means
for instance
$$(\gamma^{m_1})^\alpha{}_\beta {\hat {\cal R}}_{m_1n_1}^\beta =
\frac{1}{2} \partial_{n_1} {\psi}^\alpha\,,$$
$$(\gamma^{m_1})^\alpha{}_\beta {\hat{\cal R}}_{m_1n_1, m_2n_2}^\beta =
\frac{1}{4} \partial_{n_1} {\hat{\cal R}}_{m_2n_2}^\alpha\,,$$
etc. As a consequence, the fermionic spin $s$ tensor ${\hat{\cal
R}}^\a_{_{{[2]_1} \cdots\,[2]_{s-{1\over 2}}}}(x )$ decomposes as
follows
\begin{equation}\label{R+dRf}
{\hat{\cal R}}_{_{{[2]_1} \cdots\,[2]_{s-{1\over 2}}}}(x ) =
{{\cal R}}_{{{[2]_1} \cdots\,[2]_{s-{1\over
2}}}}(x)
+ \sum\limits_{k=1}^{s-{1\over 2}}a_k
\,\partial_{[m_1}\,\gamma_{n_1]} \ldots
\partial_{[m_k}\,\gamma_{n_k]}
{\hat{\cal
R}}_{_{{[2]_{k+1}}\cdots\,[2]_{s-{1\over 2}}}}(x )  ,
\end{equation}
where according to our notation (eqs. (\ref{D/2}) and
(\ref{D/2,121})) all the pairs of the antisymmetric indices
$[m_i\,n_i]$ are symmetrized (with weight one), and ${{\cal R}}^\a_{_{{[2]_1}
\cdots\,[2]_{s-{1\over 2}}}}(x)$ is gamma--traceless as in eq.
(\ref{gtrace}).
The coefficients $a_k$ can be determined iteratively using eq
(\ref{gammatrace}): $a_{2k-1}=  2k a_{2k}$,
 $a_{2k} = 2 (s- k + 1/2) a_{2k+1},$    $a_1=- \frac{1}{2(s - 1/2)}$.
Therefore the gamma--trace parts of the higher
rank spin--tensors do not describe any independent physical higher
spin degrees of freedom.

Eq. (\ref{trace}) implies that starting with spin 2 the trace of
the bosonic spin $s$ tensor is proportional to the second
space--time derivative of the spin $s-2$ tensor, {\it e.g.}
$$ {\hat R}_{m_1 n_1,}{}^{n_1}{}_{n_2}
= \frac{1}{8}\partial_{m_1} \partial_{n_2} \phi\,,\quad
 {\hat R}_{m_1 n_1,}{}^{n_1}{}_{n_2, m_3 n_3}
= \frac{1}{24} \partial_{m_1} \partial_{n_2}
F_{m_3 n_3}
$$
and so on for all higher spins.
Analogously to
 the case of the half-integer higher spin fields one can extract the traceless
part of the curvature decomposing ${\hat R}_{_{[2]_1 \cdots\,[2]_s
}}$ in the following way
\begin{eqnarray}\label{R+ddR}
{\hat R}_{_{[2]_1 \cdots\,[2]_s }}(x)&=&{R}_{_{[2]_1 \cdots\,[2]_s}}(x)
- \frac{1}{2s}
\partial_{[m_1}\eta_{n_1][n_2}\partial_{m_2]}
{\hat R}_{_{[2]_{3}
\cdots\,[2]_s }}(x)+ \\ \nonumber
&&+ \sum\limits_{k=2}^{[\frac{s}{2}]} b_k
\partial_{[m_1}\eta_{n_1][n_2}\partial_{m_2]}
\ldots
\partial_{[m_{2k-1}}\eta_{n_{2k-1}][n_{2k}}\partial_{m_{2k}]}
{\hat R}_{_{[2]_{2k+1}
\cdots\,[2]_s }}(x)\,,
\end{eqnarray}
where ${R}_{_{[2]_1 \cdots\,[2]_s }}(x)$ is traceless as in eq.
(\ref{tracel}), and all the pairs of the antisymmetric indices are
symmetrized (with unit weight). The exact values of the
coefficients $b_k$, which can be determined iteratively
 as in the case of the half integer higher spins, are not important for further analysis.
 The structure of (\ref{R+ddR}) tells us that the traces of the
higher rank tensors do not describe independent higher spin
degrees of freedom.

Eqs. (\ref{Bianchi1}),  (\ref{Bianchi2}) and
(\ref{transversality}) require that the tensor fields ${\hat
R}_{_{[2]_1 \cdots\,[2]_s }}(x)$ and ${\hat{\cal R}}^\a_{_{{[2]_1}
\cdots\,[2]_{s-{1\over 2}}}}(x )$  as well as the (gamma--)
traceless fields ${R}_{_{[2]_1 \cdots\,[2]_s }}(x)$ and ${{\cal
R}}^\a_{_{{[2]_1} \cdots\,[2]_{s-{1\over 2}}}}(x)$ satisfy the
 Bianchi identities,  eqs. (\ref{i}) and (\ref{closed}), and that
they are  co--closed (\ref{trans}). Thus, in accordance with the
general discussion of Sec. \ref{conformalHS}, the traceless
${R}_{_{[2]_1 \cdots\,[2]_s }}(x)$   and gamma--traceless ${{\cal
R}}^\a_{_{{[2]_1} \cdots\,[2]_{s-{1\over 2}}}}(x)$ tensor fields
are the curvatures of the higher spin gauge field potentials
satisfying `geometric' equations of motion. For instance $F_{mn}$
is the on--shell Maxwell field strength and $R_{m_1n_1,\,m_2n_2}$
is the linearized on--shell Riemann curvature.

We have thus reviewed how the free geometric equations for the
infinite set of higher spin field strengths in $D=4$ space--time
arise from the simple scalar (\ref{b}) and spinor (\ref{f}) field
equations  in $n=4$ tensorial space. The $OSp(1|2n)=OSp(1|8)$
invariance of the tensorial equations implies the $OSp(1|8)$
generalized superconformal invariance of the infinite system of the
geometric integer and half integer spin equations. Each physical
field of spin $s$ appears in this infinite spectrum only once.

\subsection{n=16, D=10}
We now turn to the more complicated case of the derivation of the
conformal higher spin geometrical equations in 10--dimensional
space from the $n=16$ tensorial space equations (\ref{b}) and
(\ref{f}).

\subsubsection{Coordinates}

In this case, the twistor--like variable $\l_\a$ is a 16--component
Majorana-Weyl spinor. The gamma--matrices
$\gamma_m^{\a\b}$ and $\gamma_{m_1\cdots m_5}^{\a\b}$ form a basis
of the symmetric $16\times 16$ matrices, so the $n=16$ tensorial
manifold is parametrized by the coordinates
\begin{equation}
X^{\a\b}= {1\over 16}\,\Big(\,x^m\gamma_m^{\a\b}+{1\over {2\cdot
5!}} \,y^{m_1\ldots m_5}\gamma_{m_1\ldots
m_5}^{\a\b}\Big)=X^{\b\a}\,, \quad (m=0,1,\ldots,9\,; \quad
\a,\b=1,2,\ldots,16)\,,
\end{equation}
where $$x^m=X^{\a\b}\gamma^m_{\a\b}\,$$ are associated with the
coordinates of the $D=10$ space--time, while the anti--self--dual
coordinates
$$y^{m_1\ldots m_5}=X^{\a\b}\gamma^{m_1\ldots m_5}_{\a\b}=-{1
\over 5!}\,\epsilon^{m_1\ldots m_5n_1\ldots n_5}y_{n_1\ldots
n_5}\,,$$ describe spin degrees of freedom. The derivative with
respect to $X^{\a\b}$ is therefore given by
\begin{equation}
\partial_{\a\b}=\gamma^m_{\a\b}\,\partial_m+
\gamma^{m_1\ldots m_5}_{\a\b}\,\partial_{m_1\ldots m_5}
=\gamma^m_{\a\b}\,\partial_m+
\gamma^{[5]}_{\a\b}\,\partial_{[5]}\,,
\end{equation}
where $\partial_{m_1\ldots m_5}\equiv \partial_{[5]}$ is the
derivative with respect to $y^{m_1\ldots m_5}\equiv y^{[5]}$,
which because of self--duality has the following property
\begin{equation}\label{d5}
\partial_{m_1\ldots m_5}\,y^{n_1\ldots n_5}
={1\over 2}\left(\delta^{n_1\cdots
\,n_5}_{m_1\cdots\,m_5}+{1\over{5!}}\,\epsilon_{m_1\ldots
m_5}{}^{n_1\ldots n_5}\right)
\end{equation}
where $ \d^{n_1\cdots \,n_i}_{m_1\cdots\,m_i}\equiv
\d^{[n_1}_{m_1}\,\d^{n_2}_{m_2}\cdots\, \d^{n_i]}_{m_i}$.

\subsubsection{Field equations}

The matrices $\gamma_{m_1m_2 m_3}^{\b\g}$ form a basis of the
 antisymmetric $16\times 16$ matrices, therefore the $n=16$ spinor
equation (\ref{f}) is equivalent to the equation
\begin{equation} (\g^n\partial_n+\g^{n_1\ldots
n_5}\partial_{n_1\ldots n_5})\g_{m_1m_2 m_3}f(x,y)=0\,.\label{f10}
\end{equation}
  By virtue of the gamma matrix properties (\ref{g1pg1}), the
multiplication of (\ref{f10}) by $\g^{m_1m_2m_3}$ leads to the
Dirac equation
\begin{equation} \g^m\partial_m
f(x,y)=0\,.
\label{Weyl10}
\end{equation}
Now taking into account (\ref{Weyl10}), using the identity
(\ref{gg}) of the Appendix and the duality relations between
$\gamma$--matrices we can rewrite eq. (\ref{f10}) in a simpler
form
\begin{equation}\label{f101}
\left(6\,\gamma_{[m_1m_2}\,\partial_{m_3]} -5!
\,\g^{n_1n_2}\partial_{m_1m_2m_3n_1n_2} -5!
\,\g^{n_1n_2n_3n_4}\,\eta_{n_4[m_1}\,\partial_{m_2m_3]n_1n_2n_3}\right)f(x,y)=0\,.
\end{equation}

 Multiplying eq. (\ref{f101}) by $\gamma_{m_4}$ and anti--symmetrizing the indices
 we get
\begin{eqnarray}
&{\hspace{-250pt}}\left(-6\,\g_{[m_1m_2m_3}\partial_{m_4]}-2\cdot
5! \,\g^{m_5}\partial_{m_1m_2m_3
m_4m_5}\right.\nonumber\\
&\left.-2\cdot 5!
\,\gamma^{n_1n_2}_{~~~~~[m_1}\,\partial_{m_2m_3m_4]
n_1n_2}-5!\,\gamma^{n_1n_2n_3}_{~~~~~~~[m_1m_2}\partial_{m_3m_4]
n_1n_2n_3}\right)f(x,y)=0\,.\label{from100}
\end{eqnarray}
We now notice that because of the self--duality of
$\gamma^{[5]\alpha\beta}$ and $\partial_{[5]}$ the last term in
(\ref{from100}) is identically zero
$$
-5!\,\gamma^{n_1n_2n_3}_{~~~~~~~[m_1m_2}\partial_{m_3m_4]
n_1n_2n_3}\,f(x,y)\equiv 0\,,
$$
so (\ref{from100}) reduces to
\begin{equation}
\left(-6\,\g_{[m_1m_2m_3}\partial_{m_4]}-2\cdot 5!
\,\g^{m_5}\partial_{m_1m_2m_3 m_4m_5}-2\cdot 5!
\,\gamma^{n_1n_2}_{~~~~~[m_1}\,\partial_{m_2m_3m_4]
n_1n_2}\right)f(x,y)= 0\,.\label{from10}
\end{equation}
As is explained below, this equation splits into
\begin{equation}
(2\,\g_{[m_1m_2m_3}\partial_{m_4]}-5!\,\g^{m_5}\partial_{m_1m_2m_3
m_4m_5})f(x,y)=0\,,\label{gammatr10}
\end{equation}
 which is an analogue of (\ref{gammatrace}), plus
\begin{eqnarray}\label{35}
\left(5\,\g_{[m_1m_2m_3}\partial_{m_4]}+
{5!}\,\gamma^{n_1n_2}_{~~~~~[m_1}\,\partial_{m_2m_3m_4]n_1n_2}
\right)\,f(x,y)=0\,.
\end{eqnarray}
Then, as a consequence of (\ref{gammatr10}) and (\ref{f101})
\begin{eqnarray}\label{51}
\left( 5\g^{n_1n_2}\partial_{m_1m_2m_3n_1n_2} + 3
\,\g^{n_1n_2n_3n_4}\,\eta_{n_4[m_1}\,\partial_{m_2m_3]n_1n_2n_3}\right)f(x,y)=0\,.
\end{eqnarray}

 A simple way to arrive at (\ref{gammatr10}) and (\ref{35})
  is to consider the general twistor--like solution
(\ref{fis}), (\ref{sollD}) of the tensorial fermionic equation
(\ref{f}). Acting on
\begin{equation}\label{Phi10}
f_\b(X)= -i\int \,d\lambda^{16}\,e^{i {1\over 32\times 5!}(\lambda
\gamma_{m_1\ldots m_5} \lambda)\, y^{m_1\ldots m_5}} \,
e^{i{1\over 16} (\lambda \gamma_m \lambda)\, x^m}
\,\lambda_\b\,\varphi(\lambda)\,+ {~~~c.c.} \,
\end{equation}
with $5!\,\g^{p_5\,\alpha\beta}\partial_{p_1p_2p_3 p_4p_5}$ we get
(up to the factor ${1\over 16}$) the following trilinear
combination of $\lambda$'s
$${1\over 2}\,(\l\g_{m_1\ldots m_5}\l)(\g^{m_5}\l)^\a.$$
Then, using the identities (\ref{usef102}), (\ref{usef106}) of the
Appendix, the basic cyclic identity (\ref{basicfier}) and taking
into account that $\l\g_{m_1m_2 m_3}\l\equiv 0$  we find that
\begin{eqnarray}
{1\over 2}(\l\g_{m_1\ldots m_4 m_5}\l)(\g^{m_5}\l)^\a &=& {1\over
2}(\l\g_{m_1\ldots
m_4}\,\g_{m_5}\l)\,(\g^{m_5}\l)^\a\label{ident101}\\
&=&-{1\over 4}(\g_{m_5} \,\g_{m_1\ldots
m_4}\l)^\a\,(\l\g^{m_5}\l)\label{ident102}\\
&=& 2(\,\g_{[m_1m_2 m_3}\l)^\a\,(\l\g_{m_4]}\l)\,.\label{ident103}
\end{eqnarray}
On the other hand, expression (\ref{ident103}) is obtained (up to
the factor ${1\over 16}$) by the action of
$2\,\g^{\a\b}_{[m_1m_2m_3}\partial_{m_4]}$ on (\ref{Phi10}). This
completes the twistor--like proof of (\ref{gammatr10}).

To prove (\ref{35}) we  multiply (\ref{ident101}) and
(\ref{ident103}) by $\gamma_{n_4}^{~~m_4}$, antisymmetrize the
indices $[m_1m_2m_3n_4]$ and use the  relations (\ref{usef102}),
(\ref{usef106}), (\ref{ident101}) and (\ref{ident103}) to get
\begin{equation}\label{3t}
{1\over 2}\,(\l\g_{m_4 m_5[m_1m_2m_3
}\l)(\g_{n_4]}^{~~~m_4m_5}\l)^\a =5\,(\,\g_{[m_1m_2
m_3}\l)^\a\,(\l\g_{n_4]}\l)\,,
\end{equation}
which is the algebraic twistor--like solution of the eq.
(\ref{35}).

The above twistor--like analysis implies  that among the equations
(\ref{f101})--(\ref{35}) and (\ref{51}) only one is independent,
while the others follow from it, provided the Dirac equation
(\ref{Weyl10}) holds (which can also be checked directly). For
instance, we can consider eqs. (\ref{Weyl10}) and
(\ref{gammatrace}) as the independent fermionic equations which
replace (\ref{f10})
\begin{eqnarray}
\g^m\partial_m f(x,y)=0\,,\label{10D}\\
(2\,\g_{[m_1m_2m_3}\partial_{m_4]}-
5!\,\g^{m_5}\partial_{m_1m_2m_3 m_4m_5})f(x,y)=0\,.\label{fermi10}
\end{eqnarray}

Analogously, the tensorial equation (\ref{b}) for the field
$b(x)$, which  in $D=10$ is equivalent to
\begin{equation}
tr[(\g^m\partial_m+\g^{m_1\ldots m_5}\partial_{m_1\ldots
m_5})\g_{p_1p_2 p_3}(\g^n\partial_n+\g^{n_1\ldots
n_5}\partial_{n_1\ldots n_5})\g_{q_1q_2
q_3}]b(x,y)=0\,,
\end{equation}
 reduces to the following set of
equations
\begin{eqnarray}
& \hspace{75pt}\partial^p\partial_p \,b(x,y)=0\,,\label{KG10}\\
{\hspace{-30pt}}
(\,\d^{[n_1n_2n_3}_{[m_1\,m_2\,m_3}\partial^{n_4]}\partial_{m_4]}
-    5!
\,\d^{[n_1}_{[m_1}\partial_{m_2}\partial_{m_3m_4]}{}^{n_2n_3n_4]}\nonumber\\
& {\hspace{-50pt}}
+5\cdot 5!\,\partial^{n_1n_2n_3n_4p}\,\partial_{m_1m_2m_3m_4p}\,)\,b(x,y)=0\,,\label{Trace10}\\
&\nonumber \\
& \hspace{45pt}
\partial^{m_5}\partial_{m_1\ldots
m_5}\,b(x,y)=0\,.\label{transversality10}
\end{eqnarray}

 Because of the self--duality of $\partial_{m_1\ldots
m_5}\,b(x,y)$ the Bianchi identity
\begin{equation}\label{second10}
\partial_{[n}\,\partial_{m_1\ldots
m_4m_5]}\,b(x,y)=0\,
\end{equation}
follows from the transversality condition
(\ref{transversality10}).

As in the case of the fermionic equations, a simple way to derive
eqs. (\ref{KG10})--(\ref{transversality10}) is to make use of the
general solution (\ref{bis}) of eq. (\ref{b}), which, in turn, is a
consequence of the general plane wave solution (\ref{sollD}) of the
$n=16$ preonic equation (\ref{l})
\begin{equation}\label{bPhi10}
 b(X)=  \int
\,d\lambda^{16}\,e^{i {1\over 32\times 5!}(\lambda
\gamma_{m_1\ldots m_5} \lambda)\, y^{m_1\ldots m_5}} \,
e^{i{1\over 16} (\lambda \gamma_m \lambda)\, x^m}
\,\varphi(\lambda)\,+ {~~~c.c.}\,.
\end{equation}
The  Klein--Gordon equation (\ref{KG10}) for $b(x,y)$ has already
been derived in this fashion (see eqs. (\ref{usef103}) and
(\ref{KGD})).

To check (\ref{transversality10}) we take the second derivative
$\partial^{m_5}\partial_{m_1\ldots m_5}$ of (\ref{bPhi10}) and
observe that it is indeed zero because of the identity
\begin{equation}\label{l5ll1l}
(\l\g_{m_1\ldots m_4 m_5}\l)(\l\g^{m_5}\l)=0\,,
\end{equation}
which in view of $\l\,\gamma_{m_1m_2m_3}\,\l\equiv 0$ follows from
(\ref{ident101})--(\ref{ident103}).

To check (\ref{Trace10}) we multiply (\ref{ident101}) and
(\ref{ident103}) by ${1\over 2\cdot 4!
}\,(\lambda\gamma^{n_1\cdots\,n_4})_\alpha$ and use the identity
(\ref{gg}) to get
$$
\hspace{-300pt}{1\over 4\cdot 4!
}\,(\lambda\gamma^{n_1\cdots\,n_4p}\l)\,(\lambda\gamma_{m_1\cdots\,m_4p}\l)
$$
$$
\hspace{100pt}={1\over 2}\,
\,\delta^{[n_1}_{[m_1}\,(\l\gamma_{m_2}\l)\,
(\l\gamma^{n_2n_3n_4]}_{~~~~~~m_3m_4]}\l)-
\,\delta^{[n_1n_2n_3}_{[m_1m_2m_3}\,
\,(\l\gamma_{m_4]}\l)\,(\l\gamma^{n_4]}\l)\,,
$$
which is the twistor--like analog of (\ref{Trace10}).

Let us now show that the equations
(\ref{fermi10})--(\ref{second10}) comprise the system of the
geometrical equations for the field strengths of the conformal
higher spin fields in ten--dimensional space--time. The expansions
of $b(x,\,y)$ and $f_\a(x,y)$ in series of $y^{[5]}\equiv
y^{m_1\cdots m_5}$ are
\begin{eqnarray}\label{ymnpqr}
b(x,\,y)&=&\phi(x)+y^{ [5]}F_{ [5]}(x) +y^{
[5]_1}\,y^{[5]_2}\,\hat R_{[5]_1
[5]_2}(x)+\sum_{s=3}^{\infty}\,y^{ [5]_1}\cdots
y^{[5]_s}\,{\hat R}_{[5]_1\, \,\cdots\, \,[5]_s}(x)\,,\nonumber\\
~&~\\
 f_\a(x,\,y)
 &=&\psi_\a(x)+y^{ [5]}{\hat{\cal R}}_{\a\,[5]}(x)+\sum_{s=5/2}^{\infty}\,y^{[5]_1}\cdots
y^{[5]_{s-1/2}}\,{\hat{\cal R}}_{\a\,[5]_1
 \cdots\, [5]_{s-1/2}}(x)\,.\nonumber
\end{eqnarray}
The scalar field $\phi(x)$ and the spinor field $\psi_\a(x)$, as
well as all the higher order tensors and spinor-tensors in
(\ref{ymnpqr}), satisfy the Klein Gordon equation (\ref{KG10}) and
hence are massless $D=10$ fields. The fermionic fields
$\psi_\a(x)$ and ${\hat{\cal R}}_{_{\a\,{[5]_1}
\cdots\,[5]_{s-{1\over 2}}}}(x )$ satisfy the Dirac equation
(\ref{10D}).

As in $D=4$,   eq. (\ref{gammatr10}) relates the gamma--trace of
the rank $5\times s-{1\over 2}$ tensor to the first derivative of the rank
$5\times (s-1)$ tensor, e.g.
$$ \gamma^{r_1}
{\hat{\cal R}}_{m_1 n_1 p_1 q_1 r_1 } ={2\over 5!} \gamma_{[m_1 n_1 p_1}
\partial_{q_1]} \psi\,.
$$
Therefore in complete analogy with $D=4$ case on can extract from
the spinor-tensor ${\hat{\cal R}}_{\a\,[5]_1 \cdots [5]_{s-{1\over
2}} }$ its gamma traceless part with the help of eq. (\ref{fermi10})
which implies that
\begin{equation}
{\hat{\cal R}}_{\a\,[5]_1 \cdots [5]_{s-{1\over 2}} } = {\cal
R}_{\a\,[5]_1 \cdots [5]_{s-{1\over 2}} } +{1\over {2\cdot
5!(s-{1\over 2}})}\,
\partial_{[m_1}\,(\gamma_{n_1 p_1q_1r_1]}\,
{\hat{\cal R}})_{\a\,[5]_2 \cdots [5]_{s-{1\over 2}} } +\ldots\,,
\end{equation}
where, according to our notation, the groups of   five
antisymmetric indices are symmetrized (with weight one), and
${\cal R}_{\a\,[5]_1 \cdots [5]_{s-{1\over 2}}}$ is
$\gamma$--traceless as in eq. (\ref{gtrace}). The dots stand for
terms proportional to higher order derivatives of lower rank
spinor--tensors.

 A novelty of the $D=10$ case with respect to $D=4$
is that eq. (\ref{Trace10}) relates the trace of the rank $5\times
s$ tensor not only to the second derivative of the rank $5\times
(s-2)$ tensor but also to the first derivative of the rank
$5\times(s-1)$ tensor, {\it e.g.}
\begin{eqnarray}
{\hat R}^{m_1 n_1 p_1 q_1 r_1,}{}_{ m_2 n_2 p_2 q_2 r_1} =
- {1\over 10 \cdot 5!}\partial^{[m_1}
\delta^{n_1}_{[m_2}\delta^{p_1}_{n_2} \delta^{q_1]}_{p_2}
\partial_{q_2]} \phi
+ {1\over 10}\partial^{[m_1} \delta^{n_1}_{[m_2} F^{p_1
q_1]}{}^{}_{n_2p_2 q_2]}\,.
\end{eqnarray}
The traceless part $R_{[5]_1\cdots\,[5]_s}$ of the tensor ${\hat{
R}}_{[5]_1\cdots\,[5]_s}$ can be extracted with the help of eq.
(\ref{Trace10}).

The traceless rank $5s$ tensor ${ R}_{[5]_1 \cdots [5]_{s} }$  is
automatically irreducible under $GL(10,\mathbb R)$ due to the
self--duality property, and it is thus associated with the
rectangular Young diagram $(s,s,s,s,s)$ which is made of five rows
of equal length $s$.   Eq. (\ref{transversality10}) is the
transversality condition, hence the rank $5s$ tensor is harmonic
(again due to self--duality) and satisfies the Bianchi identities
(\ref{second10}). In accordance with the general considerations of
Sec. \ref{conformalHS}, this implies that the traceless tensor
${R}_{[5]_1\cdots\, [5]_{s}}$ is indeed the field strength of a
chiral spin $s$ gauge field $\phi_{[4]_1\cdots \,[4]_s}$ and that
the gamma--traceless spinor--tensor ${\cal R}_{\alpha\,[5]_1\cdots\,
[5]_{s-{1\over 2}}}$ is the field strength of a fermionic chiral
spin $s$ gauge field
$\psi_{\a\,[4]_1\cdots \,[4]_{s-{1\over 2}}}$, whose symmetry
properties are described by the rectangular Young diagram
$(s,s,s,s)$ \cite{DuboisV}.

To summarize, the physical states of the  quantum $n=16$ tensorial
particle form a representation of $OSp(1|32,\mathbb R)$ which in
$D=10$ decomposes into an infinite sum containing all the chiral
integer and half--integer higher spin representations  of the
conformal group $Spin(2,10)\subset OSp(1|32,\mathbb R)$ associated
with space--time fields that satisfy the proper geometrical field
equations. Each physical field of spin $s$ appears in the spectrum
only once.

\subsection{n=8, D=6}

\subsubsection{Coordinates}

The commuting spinor $\l_\a$ is now a symplectic Majorana--Weyl
spinor (see {\it e.g.} \cite{Kugo} for details). The spinor index
can thus be decomposed as
$\a=a\otimes i$ ($\a=1,\ldots,8$; $a=1,2,3,4$; $i=1,2$). The
tensorial space coordinates $X^{\a\b}=X^{ai\,bj}$ are decomposed
into
\begin{eqnarray} && X^{ai\, bj}\, =  \,{1\over 8} \, x^m\, {\tilde\gamma}_m^{a
b} \,\epsilon^{ij}\, + {1\over 16 \cdot 3!}\,y_I^{mnp}\,
{\tilde\gamma}^{ab}_{mnp}\,\tau_I^{ij}\,,\,  \\
& & \qquad m,n,p=0,\ldots,5\,;\quad a,b=1,..., 4\,;\quad
i,j=1,2\,;\quad I=1,2,3\nonumber\end{eqnarray} where
$\epsilon^{12}=- \epsilon_{12}=1$, and $\tau^{ij}_I$ ($I=1,2,3$)
provide a basis of $2\times 2$ symmetric matrices and are
expressed through the usual $SU(2)$ group Pauli matrices, $\tau_{I
\, ij}= \e_{jj^\prime}\, \sigma_{I\, i}{}^{j^\prime}$; below we
also use $\tau_{I}^{\, ij}= \e^{ii^\prime}\, \sigma_{I\,
i^\prime}{}^j$ (see Appendix for further details). The matrices
${\tilde \gamma}^{ab}_m$ ($\gamma_{ab}^m= 1/2 \,
\varepsilon_{abcd} {\tilde \gamma}^{m\, cd}$) provide a complete
set of $4\times 4$ antisymmetric matrices with upper [lower]
indices transforming under an [anti]chiral fundamental
representation of the non--compact group $SU^*(4)\sim Spin(1,5)$.
For the space of $4\times 4$ symmetric matrices with upper [lower]
indices
  a basis is provided by  the set of self--dual
[anti--self-dual] matrices   $({\tilde \gamma}^{m n p})^{ab}$
[$\gamma_{ab}^{m n p}$],
\begin{equation}\label{ggsd}
({\tilde \gamma}^{m n p})^{ab}={1\over 3!}\epsilon^{mnpqrs}{\tilde
\gamma}^{ab}_{qrs}\; ,  \qquad \gamma_{ab}^{m n p}=- {1\over
3!}\epsilon^{mnpqrs}(\gamma_{qrs})_{ab}\; .
\end{equation}
The coordinates
\begin{equation}\label{6Dx=} x^m=\,x^{ai\,bj}\,\gamma^m_{ab}\,\epsilon_{ij}\; ,
\end{equation} are associated with $D=6$ space--time, while the
self-dual coordinates \begin{equation}\label{6Dy=asd} y_I^{mnp}=
x^{ai\,bj}\,\gamma^{mnp}_{ab}\,\tau_{I\,ij}=- {1 \over 3!
}\,\epsilon^{mnpqrs}y^I_{qrs}\; , \end{equation} describe spinning
degrees of freedom. The coefficients in (\ref{6Dx=}),
(\ref{6Dy=asd}) are chosen in such a way that the derivative with
respect to $X^{\a\b}$ is decomposed on the vector derivative
$\partial_m$ and the self--dual tensorial derivative
$\partial^I_{mnp}= {\partial \quad \over
\partial y_I^{mnp}}$ with the
unity coefficients,
\begin{equation}\label{dab=6D}
\partial_{ai\,bj}=\,\gamma^m_{ab}\,\epsilon_{ij}\,\partial_m\, + \,
\gamma^{mnp}_{ab}\,\tau^I_{ij}\,\partial^I_{mnp}\; . \qquad
\end{equation}
 The self--duality of
\begin{equation}\label{dmnp=sdd}
\partial^I_{mnp} = {1 \over
3!} \epsilon_{mnpqrs} \partial^{I\, qrs} \;
\end{equation}
implies that
\begin{equation}\label{dmnp=sd}
\tilde{\gamma}^{mnp}\,\partial^I_{mnp} = 0, \quad
\tilde{\gamma}_{[m_1}{}^{np}\,\partial^I_{m_2]np} = 0, \quad
 \tilde{\gamma}_{[m_1m_2}{}^{n}\,\partial^I_{m_2m_4]n} = 0.
\end{equation}
Note that in eq. (\ref{dab=6D})
$\gamma^{mnp}_{ab}\,\partial^I_{mnp}\not= 0$, because
$\gamma^{mnp}_{ab}$ is anti--self--dual and $\partial^I_{mnp}$ is
self--dual. Let us also notice that, as a result of the
self--duality (\ref{dmnp=sdd}) of $\partial^I_{mnp}$,
\begin{equation}\label{dydy=}
\partial^I_{mnp} \partial^{mnp\; J} = 0 \; , \qquad
\partial^I_{mnr} \partial^{J\; pqr} =
\delta_{[m}{}^{[p} \partial^{q]rs\; J} \partial^I_{n]rs} \; .
\end{equation}

Finally, let us comment on a subtlety with the `reality' condition
on the wave functions (\ref{bis}) and (\ref{fis}) and
corresponding equations of motion which one should have in mind
dealing with the $D=6$ case. The spinor $\lambda_{ai}$ is
simplectic Majorana--Weyl, but it is not real. For such a spinor
the complex conjugation condition looks as follows
\begin{equation}\label{d6mws}
({\lambda_{a}^{ i}})^*:= {\bar \lambda}_{\dot ai} =B_{\dot
a}^{~b}\epsilon_{ij}\lambda_{b}^{j}\,.
\end{equation}

The matrix $B$ is defined by the conditions
\begin{equation}\label{BgB}
B\gamma^{m}B^{-1}=(\gamma^{m})^*, \quad B\,B=-1, \quad
B^\dagger\,B=1
\end{equation}
and * and $\dagger$ denote, respectively, the complex and hermitian
conjugation.

We note that the matrix $B_{\dot a}^{~b}$ can be used to convert
the dotted indices (of the complex conjugate representation) into
undotted ones, so that one can always deal with only undotted
indices, but there is no $SU^*(4)$ invariant tensor for rising the
$SU^*(4)$ spinor indices. Thus, the spinors $\lambda^a_{ i}$ and
$\lambda_{ai}$ have different $SU^*(4)$ chiralities, one of them
is chiral (Weyl) and another one is antichiral.

The fields $b(X)$ and $f_{ai}(X)$ and their equations of motion
considered in the next subsection are self--conjugate under the
complex conjugation rules (\ref{d6mws}) and (\ref{BgB}). In this
regard we can call them `real', since as in the $D=4$ and $D=10$
cases, they lead to real integer higher spin field equations and
to simplectic Majorana--Weyl spinor equations for the
half--integer spin fields.

\subsubsection{Field equations}

The equation (\ref{f}) takes the form
\begin{equation}\partial_{ai\,bj}f_{ck}-
\partial_{ai\,ck}f_{bj}=0\,.\label{f6}\end{equation}
One can project (\ref{f6}) on the basis $\{\e , \tilde{\tau}_I \}:=
\{\e^{jk},\tau^{jk}_I\}$ ($\{\e , {\tau}_I \}:= \{\e_{jk},\tau_{I\,
jk}\}$) of complex $2\times 2$ matrices, which gives the following
system of equations (notice that $\e\e = I$, $\e\tilde{\tau}_I = -
\tau_I\e$)
\begin{eqnarray} \,\partial_m(\gamma^m_{a(b}\,  f^{}_{c)})
+ \partial^I_{mnp}(\tau_I\epsilon \, \gamma^{mnp}_{a(b}\,f^{}_{c)})&=&0\,,
\label{secondstep1bis}\\
\partial_m({\tau}_J\, \e\, \gamma^m_{a[b} \,f^{}_{c]})
- \partial^I_{mnp}(\tau_I
\tilde{\tau}_J\gamma^{mnp}_{a[b}\,f^{}_{c]})&=&0\,.\label{secondstep2bis}
\end{eqnarray}
The projection of these  equations on the basis
$\{\tilde{\g}_m^{bc},\tilde{\g}^{bc}_{mnp}\}$
of complex $4\times 4$ matrices results in
the system
\begin{eqnarray} (\gamma^m\partial_m
+ \tau_I\e\, \gamma^{mnp}\partial^I_{mnp}\,)\tilde{\g}_{qrs}f&=&0\,,
\label{thirdstep1bis}\\
({\tau}_J\e \gamma^m\partial_m -
\tau_I\tilde{\tau}_J\gamma^{mnp}\partial^I_{mnp}\,)\,\,
\tilde{\g}_q\,\,f&=&0\,,\label{thirdstep2bis}
\end{eqnarray} which is thus strictly equivalent to (\ref{f6}).
Contracting (\ref{thirdstep2bis}) with $\tilde{\g}^q$ and using
\linebreak
$\tilde{\g}^q {\g}^{n_1n_2n_3} \tilde{\g}_q \equiv 0$  one finds
that the field $f_{ai}$ obeys the Dirac equation
\begin{eqnarray} \tilde{\g}^m\partial_m
f(x,y)=0\, . \label{Weyl6}
\end{eqnarray}
Taking into account eq. (\ref{Weyl6}) and the  identities
(\ref{ggsd}) one writes eqs. (\ref{thirdstep1bis}) and
(\ref{thirdstep2bis}) as
\begin{eqnarray} && (\gamma_{[m_1m_2}\partial_{m_3]}
- 2  \, (\tau_I\e) \, \partial^I_{m_1m_2m_3}   - 3! \, (\tau_I\e)
\gamma^n{}_{[m_1}\partial^I_{m_2m_3]n})f=0 \; ,
\label{4step1bis}\\
&& ({\tau}_J\e \partial_m - 3\,
\tau_I\tilde{\tau}_J\gamma^{np}\partial^I_{mnp}\,)\,f=
0\,,\label{4step2bis}
\end{eqnarray}

Contracting  eq. (\ref{4step2bis}) with $\tau_J\e$ and using the
identities $\tau_I\tilde{\tau}_I=-3$,
$\tilde{\tau}_I\tau_J\tilde{\tau}_I=- \tilde{\tau}_J$ one finds
\begin{eqnarray}
 && (\partial_m - {\tau}_I\e \,\gamma^{np}\partial^I_{mnp}\,)\,f= 0
 \;.
 \label{5step2bis}
\end{eqnarray}
Multiplying this by $\tau_J$ and using $\tau_I\e = - \e
\tilde{\tau}_I$ one finds $(\e \tilde{\tau}_J \partial_m -
\tau_J\tilde{\tau}_I\gamma^{np}\partial^I_{mnp}\,)\,f= 0$ which,
together with (\ref{4step2bis}), implies
\begin{eqnarray}
 && {\tau}_J\tilde{\tau}_I \gamma^{np}\partial^I_{mnp}\,f= -3
\gamma^{np}\partial^J_{mnp}\,f \qquad \Leftrightarrow \qquad
{\tau}_I\tilde{\tau}_J \gamma^{np}\partial^I_{mnp}\,f=
\gamma^{np}\partial^J_{mnp}\,f \; . \qquad  \label{6step2bis}
\end{eqnarray}
The consistency of (\ref{6step2bis}) can be easily checked by
contracting (any of its forms)  with $\tilde{\tau}_J$. This is
satisfied identically. Using (\ref{6step2bis}) one finds from eq.
(\ref{4step2bis}) that
\begin{eqnarray}
(\, {\tau}_I\e  \, \partial_{m} - 3 \g^{np}\partial^I_{mnp})\,
f(x,y)=0\, . \label{0gtr6}
\end{eqnarray}
Eq. (\ref{0gtr6}) comprises the original eq. (\ref{4step2bis}) and
all its consequences. On the other hand, multiplying eq.
(\ref{0gtr6}) and using eqs. (\ref{Weyl6}) and (\ref{dmnp=sd}),
one finds
 \begin{eqnarray}
(\,{\tau}_J\e\, \tilde{\g}_{[m}\partial_{n]} +\, 3!\,
\tilde{\g}^p\partial^J_{mnp})f(x,y)=0\, .
\label{gammatr6}\end{eqnarray} This is an equivalent form of
(\ref{0gtr6})\footnote{To check this one multiplies (\ref{gammatr6})
by $\g^n$ and uses the Dirac equation (\ref{Weyl6}).} which is most
useful for the analysis below.

Now let us turn to eq. (\ref{4step1bis}). Contracting it with
${\tilde\g}^{m_3}$ and using the second identity in (\ref{dmnp=sd}) one
finds
\begin{eqnarray}
(\tilde{\g}_{[m}\partial_{n]} + 2 \tau_I\e \,
\tilde{\g}^p\partial^J_{mnp})f(x,y)=0\, ,
\label{5step1bis}\end{eqnarray} which can also be obtained
multiplying (\ref{5step2bis}) by ${\tilde\gamma}_n$ and
antisymmetrizing the indices $m$ and $n$. On the other hand,
multiplying  eq. (\ref{4step1bis}) by ${\tilde\gamma}_{m_4}$,
antisymmetrizing the indices and using the third identity in
(\ref{dmnp=sd})
 one finds
\begin{eqnarray}
(\, \tilde{\g}_{[m_1m_2m_3}\partial_{m_4]}+\, 4\, \tau_I\e
\tilde{\g}_{[m_1}\partial^I_{m_2m_3m_4]})f(x,y)=0\, .
\label{6step1bis}\end{eqnarray} This equation  is dual to
(\ref{5step1bis}). Indeed, multiplying  (\ref{6step1bis}) by
$\e^{n_1n_2m_1\ldots m_4}$ and using the self--duality of the
derivative $\partial^I_{[m_1m_2m_3]}$ and of
$\tilde{\g}_{m_1m_2m_3}:= ({\tilde\g}_{m_1m_2m_3})^{ab}$ (not to be
confused with ${\g}_{m_1m_2m_3}:= ({\g}_{m_1m_2m_3})_{ab}$ which
is anti--self--dual) one arrives at (\ref{5step1bis}). Thus one
concludes that eq. (\ref{4step1bis}) is not independent and that
eqs. (\ref{thirdstep1bis})-(\ref{thirdstep2bis}) are  equivalent
to the system of two equations (\ref{Weyl6}) and (\ref{gammatr6}),
namely
\begin{eqnarray}
\tilde{\g}^{m}\partial_{m} f(x,y)&=& 0 \; , \qquad \label{Weyl6*}
\\  (\,\tau^I\e \tilde{\g}_{[m}\partial_{n]}+\, 3!\,
\tilde{\g}^p\partial^I_{mnp})f(x,y)&=& 0\,.
\label{gammatr6*}\end{eqnarray} One more consequence of these
equations is useful
\begin{eqnarray}
({\g}_{[m_1m_2}\partial_{m_3]}+\, 4\, \tau_I\e
\partial^I_{m_1m_2m_3})f(x,y)=0\, .
\;  \qquad \label{sigmatr6*}\end{eqnarray} It can be obtained by
comparing (\ref{4step1bis}) with the result of the contraction of
 (\ref{6step1bis}) with $\g^{m_4}$. Indeed, such a comparison
 results in
$\tau_I\e{\g}^n{}_{[m_1}\partial^I_{m_2m_3]n}f = -
\tau_I\e\partial^I_{m_1m_2m_3}f$ whose substitution in
(\ref{4step1bis}) gives (\ref{sigmatr6*}).

The system of the field equations for the tensorial space scalar
$b(x,y)$ originating from the $n=8$, $D=6$ version of eq.
(\ref{b}) consists of
\begin{eqnarray}
\partial^p\partial_p \,b(x,y)&=&0\,, \label{KG6} \\
\partial^p\partial^I_{mnp}\,b(x,y)= 0\qquad \Leftrightarrow
\qquad
\partial^{}_{[m}\partial^I_{npq]}\,b(x,y)&=&0 \, , \label{transversality6} \\
\left(\,\partial^{I\,\,m_1m_2p}\,\partial^J_{n_1n_2p}
 \,- {{i \epsilon^{IJK}} \over (3!)}
\partial^{}_{[n_1} \partial_{n_2]}^{K\,\,\,\,m_1 m_2}
+ {{\delta^{IJ}} \over {(3!)}^2}\d^{[m_1}_{[n_1}\,\partial_{}^{m_2]}
\partial^{}_{n_2]}
\right)\,b(x,y)=0 & & \Leftrightarrow \qquad \nonumber \\
\Longleftrightarrow \qquad  \cases{ \left(\,\partial_{}^{m_1m_2p\,
(I}\,\partial^{J)}_{n_1n_2p} + {1\over
{(3!)}^2}\delta^{IJ}\d^{[m_1}_{[n_1}\,\partial_{}^{m_2]}
\partial^{}_{n_2]}
\right)\,b(x,y)=0 \cr \left(\,\partial_{}^{m_1m_2p\,
[I}\,\partial^{J]}_{n_1n_2p}
 \,- {{i \epsilon^{IJK}}\over (3!)}
\partial^{}_{[n_1} \partial_{n_2]}^{K\,\,\,\,m_1 m_2}
\right)\,b(x,y)=0 } \; . \hspace{-1cm} & & \label{Trace6}
\end{eqnarray}
One more useful equation is
\begin{eqnarray}
(\delta^{[n_1 n_2}_{[m_1 m_2}\partial^{}_{m_3]}\partial^{n_3]}+\, 8
\,
\partial^I_{m_1 m_2 m_3}\partial_I^{n_1 n_2
n_3})b &=& 0\, . \qquad \label{trso3}
\end{eqnarray}

A simple way to obtain eqs. (\ref{KG6}), (\ref{transversality6})
and (\ref{Trace6}) is to observe that the derivative
$\partial_{\gamma\,\delta} b$ of the bosonic field $b(x, y)$
obeying the bosonic equation (\ref{b}) can be treated as a set of
solutions of the fermionic equation (\ref{f}) (the extra spinor
index $\delta$ being regarded as the label of the fermion--like
solutions). Then in view of the form of eqs. (\ref{Weyl6*}) and
(\ref{gammatr6*}) on finds that the independent bosonic equations
following from (\ref{b}) are
\begin{eqnarray}
\tilde{\g}^{m}\partial_{m}\, (\epsilon\cdot \gamma^n\,\partial_n\, +
\, \tau_I\cdot \gamma^{npq}\,\partial^I_{npq}) b(x,y)&=& 0 \; ,
\qquad \label{bWeyl6*}
\\  (\,\tau_I\e \cdot \tilde{\g}_{[m_1}\partial_{m_2]}+\, 3!\,
\tilde{\g}^{m_3}\partial^I_{m_1m_2m_3})\, (\epsilon\cdot
\gamma^n\,\partial_n\, + \, \tau_J\cdot
\gamma^{npq}\,\partial^J_{npq}) b(x,y)&=& 0\, . \label{bgammatr6*}
\end{eqnarray}
Now obsereve that the terms in (\ref{bWeyl6*}) proportional to
$\e_{ij}$ and $\tau_{I\, ij}$  should vanish separately. They
produce, respectively, the Klein--Gordon equation (\ref{KG6}) and
eq. (\ref{transversality6}) (to derive the latter one should
remember that $\tilde{\gamma}^{m_1\ldots m_4}= 1/2 \e^{m_1\ldots
m_4pq}\tilde{\gamma}_{pq}$; not to be confused with
 ${\gamma}^{m_1\ldots m_4}= (\tilde{\gamma}^{m_1\ldots m_4})^T=  - 1/2
 \e^{m_1\ldots
m_4pq}{\gamma}_{pq}$ where the sign is opposite). With this in
mind we find that the only independent part of eq.
(\ref{bgammatr6*}) is
\begin{eqnarray}
\mathrm{tr} \left[\tau_J\e\cdot \gamma_{n_1n_2} (\,\tau_I\e \cdot
\tilde{\g}_{[m_1}\partial_{m_2]}+\, 3!\,
\tilde{\g}^{m_3}\partial^I_{m_1m_2m_3})\, (\epsilon\cdot
\gamma^n\,\partial_n\, + \, \tau_K\cdot
\gamma^{npq}\,\partial^K_{npq})\right]\;  b(x,y)&=& 0\, , \;
\nonumber \\ \label{bgammatr6**}
\end{eqnarray}
which gives eq. (\ref{Trace6}).

Alternatively, as in $D=4$ and $D=10$, instead of direct
computations one can obtain the field equations of $b(X)$ and
$f_\alpha(X)$ by using the plane wave solution (\ref{soll})
$$ \Phi (x, \lambda) = e^{i{1\over 16\cdot 3! } (\lambda {\tilde\gamma}_{mnp}\tau^I
\lambda)\, y_I^{mnp}} \, e^{i {1\over 8}(\lambda
{\tilde\gamma}_m\epsilon \lambda)\, x^m} \phi(\lambda)$$ and Fierz
identities for the $D=6$
$\gamma$--matrices.

To analyse the consequences of eqs. (\ref{Weyl6*})--(\ref{trso3})
in the effective $D=6$ higher spin field theory we expand $b(X)$
and $f_\alpha(X)$ in series of $y^{ [3]}_I$
\begin{eqnarray}\label{ymnpqr6}
b(x,\,y)&=&\phi(x)+y^{ [3]}_IF_{ [3]}^I(x) +y^{
[3]_1}_{I_1}\,y^{[3]_2}_{I_2}\,\hat R_{[3]_1
[3]_2}^{I_1 I_2}(x)+\sum_{s=3}^{\infty}\,y^{ [3]_1}_{I_1}\cdots
y^{[3]_s}_{I_s}\,{\hat R}^{I_1 \cdots I_s}_{[3]_1\,
\,\cdots\, \,[3]_s}(x)\,,\nonumber\\
~&~\\
 f_\a(x,\,y)
 &=&\psi_\a(x)+y^{ [3]}_I{\hat{\cal R}}^{I}_{\a\,[3]}(x)
+\sum_{s=5/2}^{\infty}\,y^{[3]_1}_{I_1}\cdots y^{[3]_{s-1/2}}_{I_{s-
1/2}}\,{\hat{\cal R}}^{ I_1 \cdots I_{s- 1/2}}_{\a \,[3]_1
 \cdots\, [3]_{s-1/2}}(x)\,.\nonumber
\end{eqnarray}

 As in $D=4$, and 10, the equation
(\ref{gammatr6}) relates the gamma--trace of the spin $s$
spinor--tensor to the first derivative of the spin $s-1$
spinor--tensor, {\it e.g.}
\begin{equation}
{\tilde\gamma}^p {\hat {\cal R}}_{mnp}^I = - {1\over 3!}
\partial_{[m}{\tilde\gamma}_{n]} \tau^I \psi\,.
\end{equation}
A novelty of the $D=6$ case is that in virtue of eq.
(\ref{sigmatr6*})also the `tau-trace' of the spin $s$
spinor--tensor is related to the first derivative of the spin
$s-1$ spinor--tensor, {\it e.g.}
\begin{equation}
\tau_I \e {\hat {\cal R}}_{mnp}^I = - {1\over 4}
\partial_{[m}\gamma_{np]}  \psi\,.
\end{equation}
Thus the gamma--traces and the tau--traces of the spinor--tensors
do not correspond to independent physical degrees of freedom.

The equation (\ref{Trace6}) relates the trace of the spin $s$
field strength to the first derivative of the spin $s-1$  and to
the second derivative of the spin $s-2$ field strength. For
instance,
\begin{equation}
{\hat R}^{I_1I_2}_{m_1 n_1 p,m_2 n_2}{}^{p} = \frac{i}{12}
\epsilon^{I_1I_2I_3}
\partial^{}_{[m_2} F^{I_3}_{n_2]m_1 n_1} -
\frac{1}{2 {(3!)}^2}\delta^{I_1I_2}\partial_{[m_1}\eta_{n_1][m_2}
 \partial_{n_2]}  \phi\,.
\end{equation}
Eq. (\ref{trso3}) relates the $SO(3)$ trace of the
spin--$s$ field strength to the second derivatives of the spin
$s-2$ field strength, for example
\begin{equation} \delta_{I_1 I_2}
{\hat R}^{I_1 I_2}_{m_1 n_1 p_1,}{}^{ m_1 n_2 p_2} =- \frac{1}{16}
\delta^{[m_2 n_2}_{[m_1
n_1}\partial^{}_{p_1]}\partial^{p_2]}\phi\,.
\end{equation}

The transversality of ${\hat R}^{I_1 \cdots I_s}_{[3]_1\, \,\cdots\,
\,[3]_s}(x)$ and ${\hat{\cal R}}^{ I_1 \cdots I_{s- 1/2}}_{\a
\,[3]_1 \cdots\, [3]_{s-1/2}}(x)$ is the consequence of eq.
(\ref{transversality6}). Their self--duality in each set of
antisymmetric indices is automatic. The (gamma) traceless parts
${R}^{I_1 \cdots I_s}_{[3]_1\, \,\cdots\, \,[3]_s}(x)$ and ${{\cal
R}}^{ I_1 \cdots I_{s- 1/2}}_{\a \,[3]_1
 \cdots\, [3]_{s-1/2}}(x)$ of the (spinor)--tensors describe
 propagating higher spin degrees of freedom
corresponding to $Spin(1,5)$ irreps
 characterized by rectangular Young diagrams with three
rows of equal length. They are the field strengths of the gauge
fields characterized by rectangular Young diagrams with two rows.
All this implies that the propagating fields carry irreps of the
conformal group $Spin(2,6)$.

A new feature of the $D=6$ case is the degeneracy of these irreps
due to the internal $SO(3)$ symmetry. The $GL(6,\mathbb R)$
irreducibility implies symmetry under the exchange of
multi--indices. This property, along with the commutativity of
$y_I^{[3]}$, implies that the field strengths ${R}^{I_1\ldots
I_s}_{[3]_1\cdots [3]_s}$ and  ${{\cal R}}^{I_1 \cdots I_{s-
1/2}}_{\a \, [3]_1
 \cdots\, [3]_{s-1/2}}(x)$ are also symmetric in the
internal $SO(3)$ indices $I$. This leads to the fact that each
propagating field corresponds to a spin--$s$ irrep of $SO(3)$, thus
the degeneracy of the spin--$s$ irrep is equal to $2[s]+1$.

In other words, the quantum spectrum of the tensorial $n=8$
superparticle is formed by an infinite number of conformally
invariant (self-dual) ``multi--3--form" higher spin fields in $D=6$
whose number for each value of spin $s$ is $2[s]+1$, and which form
the $(2[s]+1)$-dimensional representation of the group $SO(3)$. This
differs from the cases of $n=4,~D=4$ and $n=16, ~D=10$, where the
conformal fields of each spin $s$ appear in the quantum spectrum
only once.

\setcounter{equation}0\section{Conclusion}

In this paper we have analyzed the geometrical structure of
conformally invariant higher spin fields and have shown that in
$D=3,4,6$ and $10$ space--time the massless conformal higher spin
fields arise as a result of the quantization of the dynamics of the
twistor--like particle, respectively, in $n=2,4,8,$ and $16$
tensorial space. The $D=3$ and $D=4$ cases have already been
considered in the literature, while the $D=6$ and $D=10$ results are
new.

In each of these cases, the infinite sum of irreps of the
conformal group $Spin(2,D)$ gets combined into an
infinite--dimensional representation of the supergroup
$OSp(1|2n,{\mathbb R})$ (with $n=2(D-2)$). The latter is
associated with the solutions of the $OSp(1|2n,{\mathbb
R})$--invariant scalar and spinor field equation in  tensorial
space. The superfield form of these equations, both in flat
tensorial superspace and on the supergroup manifold
$OSp(1|n,{\mathbb R})$ was constructed in \cite{Bandos:2004nn}.
When reduced to   the effective $D=4$, 6 and 10 space--time the
tensorial equations give rise, in a very natural way, to geometric
conformal higher spin field equations in the Bargmann--Wigner form.

  To conclude let us discuss possible directions in which
present work might be developed:
\begin{itemize}

\item One of them is to generalize above results
to the $AdS_D$ space whose tensorial extension is the group
manifold $Sp(n,{\mathbb R})$
\cite{blps,Misha,Plyushchay:2003gv,Plyushchay:2003tj}. For
example, one may take as the starting point the wave function
\cite{Plyushchay:2003tj}
\begin{equation}\label{ls} \Phi(X^{\a\b},\l) =\int\, d^ny\,
\sqrt{\det G^{-1}(X)}\, e^{{i}X^{\a\b}(\l_\a+ {1\over {8r}}y_\a
)(\l_\b+{1\over {8r}}y_\b) +i \l_\a
y^\a}\,\widetilde{\varphi}(y)\,,
\end{equation}
where $G^{-1\b}_\a(X)=\delta^\b_\a+{1\over {4r}} X^{~\b}_\a$, and
$r$ is the AdS radius, and derive the $Sp(n,\mathbb R)$ analog of
the field equations (\ref{b})--(\ref{f}) \cite{Plyushchay:2003tj}
\begin{eqnarray}\label{bsp}
\nabla_{\a[\b}\nabla_{\g]\d}b&=&{1\over
{16r}}\left(C_{\a[\b}\nabla_{\g]\d}- C_{\d[\g}\nabla_{\b]\a} +
2C_{\b\g}\nabla_{\a\d}\right)b\\&&+{1\over
{64r^2}}\left(2C_{\a\d}C_{\b\g}-C_{\a[\b}C_{\g]\d}\right)b\,,
\end{eqnarray}
\begin{equation}\label{fsp}
\nabla_{\a[\b}f_{\g]}=-{1\over
{4r}}\left(C_{\a[\g}f_{\b]}+2C_{\b\g}f_\a\right)\,,
\end{equation}
where $\nabla_{\a\b}\equiv
G_\a^{-1\g}(X)G_\b^{-1\d}(X)\partial_{\g\d}$. However, to reduce the
tensorial $Sp(n,{\mathbb R})$ model to the higher spin field theory
on $AdS_D$ by disentangling   the $x^m$ and
$y^{mn\cdots q}$ dependence is much more cumbersome problem than
in flat tensorial space due to the complicated $X^{\a\b}$
dependence of the plane wave solution $\Phi(X,\l)$ and of the
covariant derivative $\nabla_{\a\b}$. To this end one may also try
to use other realizations of the $Sp(n,{\mathbb R})$ model
considered in \cite{Misha,Plyushchay:2003gv}, or its twistor
counterpart constructed in \cite{Martin}.

\item It would be also interesting to obtain the $n=8,~D=6$ spectrum by
expanding the wave function in $\l_\alpha$. The two--component
quaternionic formalism can be useful for this purpose
\cite{Kugo,bem}, like the Weyl spinor formalism for the case
$n=4,~D=4$. This can provide a new realization of the
$OSp(1|16,\mathbb R)$ infinite--dimensional irreducible representations.

\item The reduction of the $D=6$ model to $D=5$
produces an infinite tower of completely symmetric gauge fields of
all spins with exactly identical internal $SO(3)$ structure for a
given spin $s$, as can be easily seen. In analogy with Hull's
conjecture \cite{Hull:2000rr}, a strong coupling limit of a
hypothetical $D=5$ interacting higher-spin theory might be expected
to be an interacting $D=6$ exotic theory whose free limit is the
$n=8$ tensorial model considered in this paper. Though appealing,
such a scenario seems to be difficult to realize. Indeed, switching
on interactions is still a challenging open problem for gauge fields
which are either higher-spin, chiral, or of mixed
symmetry\footnote{Some no--go theorems have recently been proved for
chiral form  and two--column field self--interactions (see,
respectively, \cite{BHS} and \cite{BBC}, and refs. therein).}
(especially when they possess \textit{all} these properties
simultaneously). Note that $AdS_5$ and $AdS_7$ twistor counterparts
of the tensorial model have recently been discussed in
\cite{Martin}.

\item A possibility of introducing non--linearity in higher spin field
equations directly in a curved tensorial superspace was analyzed
in \cite{Bandos:2004nn}. It was shown that if one tries to
maintain manifest conformal invariance also in the non--linear
theory this puts too severe restrictions on the geometry of the
tensorial superspace and does not lead to higher spin
interactions. This indicates that a possible way out might be
related to breaking conformal symmetry.

\end{itemize}

{\bf Acknowledgments}. We would like to thank E. Bergshoeff, N.
Boulanger, O. Chandia, J. Gomis, K. Lechner, A. Nurmagambetov, J.-H.
Park, P. Pasti, M. Tonin, P. Townsend and M. Vasiliev for useful
discussions. This work was partially supported by the research
grants BFM2002-03681 from the Ministerio de Educaci\'on y Ciencia
and from EU FEDER funds, Grupos 03/124 from the Generalitat
Valenciana, N 383 of the Ukrainian State Fund for Fundamental
Research, the INTAS Research Project N 2000-254, the EU
MRTN-CT-2004-005104 grant `Forces Universe', and by the MIUR
contract no. 2003023852.

\def\theequation{A.\arabic{equation}}
\setcounter{equation}0
\section*{Appendix. Useful gamma matrix identities}

All antisymmetrizations of indices are denoted by brackets
$[\,\,\,]$ and have unit weight. All symmetrizations of indices are denoted by brackets
$(\,\,\,)$ and have unit weight.
Some of the  identities presented here are taken from
\cite{KennedyVanProey}.

\subsection*{Any dimension D}

The Clifford algebra is $$\g_m\g_n+\g_n\g_m=2\,\eta_{mn}\,.$$
The matrices $$\gamma_{m_1\ldots m_p}\equiv
\g_{[m_1}\ldots\g_{m_p]}$$ satisfy the orthonormality relations
\begin{equation}
tr[\g_{m_1\ldots m_p}\g^{n_1\ldots n_q}]=
(-)^{p(p-1)/2}\,\,p!\,\,\d_{m_1\ldots m_p}^{n_1\ldots\,
n_q}\,\d_{pq}\,tr[1]\,,\label{orthon}
\end{equation}
where the Kronecker symbols are defined by
\begin{equation}\d_{m_1\ldots m_p}^{n_1\ldots\, n_p}\equiv
\d^{[n_1}_{m_1}\ldots\d^{n_p]}_{m_p}=\d^{n_1}_{[m_1}\ldots\d^{n_p}_{m_p]}\,.
\end{equation}

\begin{eqnarray}
\gamma_{m_1\ldots m_{p+1}} &=& \gamma_{m_1\ldots m_p}\g_{m_{p+1}}\,
- \,p\,\,\gamma_{[m_1 \ldots
m_{p-1}}\,\eta_{m_p]m_{p+1}},\label{usef102}\\
\g_n\gamma_{m_1\ldots m_p}&=&(-)^p\,\gamma_{m_1\ldots m_p}\g_n\,
+\,2p\,\,\eta_{n[m_1}\gamma_{m_2 \ldots m_p]}\,,\label{usef106}
\end{eqnarray}
or in general
\begin{equation}\label{gg}
\gamma^{n_1\cdots n_i}\,\gamma_{m_1\cdots
m_j}=\sum_{k=0}^{k=min(i,j)}\,{{i!\,j!}\over{(i-k)!\,(j-k)!\,k!}}\,\gamma^{[n_1\cdots
n_{i-k}}_{~~~~~~~~~~[m_{k+1}\cdots
m_j}\delta^{n_i}_{m_1}\,\delta^{n_{i-1}}_{m_2}\cdots\,\delta^{n_{n-k+1}]}_{m_{k}]}\,.
\end{equation}

 We also use the following identity
\begin{eqnarray}\label{g1pg1}
\gamma^m \, \gamma^{n_1\cdots n_q}\,\gamma_{m}= (-)^q \, (D-q)  \,
\gamma^{n_1\cdots n_q} \; , \qquad \nonumber \\
 \gamma^{n_1n_2 n_3}\, \gamma^m \, \gamma_{n_1n_2 n_3} = -
 (D-6)\, (D-1)\, (D-2)\, \gamma^m \; .
\end{eqnarray}

\subsection*{D = 3, 4, 6, 10}

These dimensions are respectively associated to the division
algebras $\mathbb R$, $\mathbb C$, $\mathbb H$, and $\mathbb O$
\cite{Kugo}. A common property of the gamma matrices considered in
this paper is the Fierz identity
\begin{equation} \label{D101}
(\gamma^m)_{\alpha (\beta} (\gamma_m)_{\gamma \delta)} =0\,.
\end{equation}

\subsection*{D = 6}

\begin{equation} \label{D601}
\gamma^m_{ab}\tilde{\gamma}_m^{cd} =  - 4 \delta_{[a}{}^{c}
\delta_{b]}{}^{d} \; ,  \qquad  \gamma^m_{ab}= {1\over 2}
\epsilon_{abcd}  \tilde{\gamma}^{m\, cd} \; ,
\end{equation}
\begin{equation} \label{D61}
(\gamma_m)_{ab} (\gamma^m)_{cd}= -2\epsilon_{abcd}\,,
\end{equation}
from which originates
\begin{equation} \label{D62}
(\gamma_m)_{a(b} (\gamma^m)_{c)d}=0
\end{equation}
The matrices $\tau_I:= \tau_{I \, ij}= \e_{jj^\prime}\, \sigma_{I\,
i}{}^{j^\prime}$ and $\tilde{\tau}_I:= \tau_{I}^{\, ij}=
\e^{ii^\prime}\, \sigma_{I\, i^\prime}{}^j$ ($I=1,2,3$) obey
\begin{eqnarray}\label{tt=}
\tilde{\tau}_I \, {\tau}_J + \tilde{\tau}_J \, {\tau}_I = - 2
\delta_{IJ}\quad \Leftrightarrow \quad {\tau}^{ij^\prime}_I \,
{\tau}_{Jj^\prime j} + {\tau}^{ij^\prime}_J \, {\tau}_{Ij^\prime
j}  = - 2 \delta_{IJ} \delta_j{}^i\; . \\
\label{Pauli1} \tau_I^{ii^\prime}\tau_{I\, jj^\prime}=- 2
 \delta_{[j}{}^i  \delta_{j^\prime]}{}^{i^\prime}
\qquad \Leftrightarrow \qquad \tau_{I\, ii^\prime}\tau_{I
jj^\prime}= 2 \e_{i(j}\e_{j^\prime)i^\prime} \; .
\end{eqnarray}
 The appearance of Pauli matrices $\sigma_I:=  \sigma_{I\, i}{}^j $
($I=1,2,3$) is reminiscent of the $D=6$ quaternionic structure
$$\sigma_I\, \sigma_J = \delta_{IJ} + i \e_{IJK} \sigma_K\; . $$
The antisymmetric spin--tensor  $\e_{ij}$ and its inverse
$\e^{ij}$ are used to lower and to rise isospinorial $SU(2)$
indices, see above and
$$
\e^{ik} \e_{k j} = \d_j^i \; , \qquad \tau_{I\, ij}=
\e_{ii^\prime}\e_{jj^\prime} \tau_{I}^{ i^\prime j^\prime}
$$

\subsection*{D = 10}

The set of $16\times 16$ symmetric matrices with respect to the pair of lower
indices $\alpha\beta$ is given
by $\gamma^m_{\alpha \beta}$ and by
$\gamma^{m_1 \ldots m_5}_{\alpha \beta}$ which are self--dual in
spacetime indices $$\g^{m_1\ldots m_5}_{\alpha \beta}={1 \over
5!}\,\epsilon^{m_1\ldots m_5n_1\ldots n_5} \g_{n_1\ldots
n_5}{}_{\alpha \beta}\,.$$ In contrast, $\gamma^{m_1 \ldots m_5\,
\alpha \beta}$,  are anti--self--dual
$$\g^{m_1\ldots m_5\; \alpha \beta}= - {1 \over
5!}\,\epsilon^{m_1\ldots m_5n_1\ldots n_5} \g_{n_1\ldots n_5}{}^{
\alpha \beta}\,.$$
As a result
$$ \mathrm{tr} (\g^{m_1\ldots m_5} \,
\tilde{\g}_{n_1\ldots n_5}) := \g^{m_1\ldots m_5}_{\a\b}
{\g}^{\b\a}_{n_1\ldots n_5}
 = 16 \cdot 5! \left( \delta_{[m_1}{}^{[n_1} \ldots
\delta_{m_5]}{}^{m_5]} \; + {1 \over 5!}\ \epsilon_{m_1\ldots
m_5}{}^{n_1\ldots n_5} \right) \; ,
$$
 in distinction to $D\not= 10$  (for $D>5$) where only
 the first term is present.


\begin{thebibliography}{99}

\bibitem{Bandos:1998vz}
I.~A.~Bandos and J.~Lukierski,
Mod.\ Phys.\ Lett.\ A {\bf 14}, 1257 (1999) {\tt hep-th/9811022}.

\bibitem{Bandos:1999qf}
I.~A.~Bandos, J.~Lukierski and D.~P.~Sorokin,
Phys.\ Rev.\ D {\bf 61}, 045002 (2000) {\tt hep-th/9904109}.

\bibitem{fronsdal1}
C. Fronsdal, {\it Massless particles, ortosymplectic symmetry and
another type of Kaluza--Klein theory}, Preprint UCLA/85/TEP/10, in
``Essays on Supersymmetry", Reidel, 1986 (Mathematical Physics
Studies, v. 8).
\bibitem{Vasiliev:2001zy}
M.~A.~Vasiliev,
Phys.\ Rev.\ D {\bf 66}, 066006 (2002) {\tt hep-th/0106149}.
\\
M.~A.~Vasiliev, {\it Relativity, causality, locality, quantization
and duality in the Sp(2M) invariant generalized space-time,} {\tt
hep-th/0111119}.

\bibitem{Misha}
V.E. Didenko and M.A. Vasiliev, {\it Free field dynamics in the
generalized $AdS$ (super)space}, {\tt hep-th/0301054}.

\bibitem{Plyushchay:2003gv}
M.~Plyushchay, D.~Sorokin and M.~Tsulaia,
JHEP {\bf 0304}, 013 (2003) {\tt hep-th/0301067}.

\bibitem{Plyushchay:2003tj}
M.~Plyushchay, D.~Sorokin and M.~Tsulaia,
{\it GL flatness of OSp(1$|$2n) and higher spin field theory from
dynamics in tensorial spaces,} {\tt hep-th/0310297}.


\bibitem{V01}
M.A. Vasiliev, {\it Progress in higher spin gauge theories}, {\tt
hep-th/0104246}, {\it Higher-spin theories and $Sp(2M)$ invariant
space--time}, {\tt hep-th/0301235}, and refs. therein.

\bibitem{Misha03-04b}
M.~A.~Vasiliev,
Phys.\ Lett.\ B {\bf 567} (2003) 139, {\tt hep-th/0304049};
Fortsch.\ Phys.\  {\bf 52} (2004) 702,  {\tt hep-th/0401177}.
\\
M.~A.~Vasiliev and V.~N.~Zaikin,
Phys.\ Lett.\ B {\bf 587}, 225 (2004) {\tt hep-th/0312244}.


\bibitem{Bandos:2003us}
I.~A.~Bandos, J.~A.~de Azc\'arraga, M.~Pic\'on and O.~Varela,
Phys.\ Rev.\ D {\bf 69} (2004) 085007, {\tt hep-th/0307106}.
\\
I.~A.~Bandos, J.~A.~de Azc\'arraga, J.~M.~Izquierdo, M.~Pic\'on
and O.~Varela,
Phys.\ Rev.\ D {\bf 69} (2004) 105010, {\tt hep-th/0312266}.

\bibitem{blps}
I.~A.~Bandos, J.~Lukierski, C.~Preitschopf and D.~P.~Sorokin,
Phys.\ Rev.\ D {\bf 61} (2000) 065009 {\tt hep-th/9907113}.

\bibitem{othertopics}
M.~A.~Vasiliev,
Russ.\ Phys.\ J.\  {\bf 45}, 670 (2002) [Izv.\ Vuz.\ Fiz.\  {\bf
2002N7}, 23 (2002)] {\tt hep-th/0204167}.

\bibitem{Gelfond:2003vh}
O.~A.~Gelfond and M.~A.~Vasiliev, {\it Higher rank conformal
fields in the Sp(2M) symmetric generalized space-time,} {\tt
hep-th/0304020}.



\bibitem{deWit:1979pe}
B.~de Wit and D.~Z.~Freedman,
Phys.\ Rev.\ D {\bf 21}, 358 (1980).
\bibitem{Dirac}
P.~A.~M.~Dirac,
Proc.\ Roy.\ Soc.\ Lond.\  {\bf 155A} (1936) 447.
\bibitem{Bargmann}
V.~Bargmann, E.~P.~Wigner,
Proc.\ Nat.\ Acad.\ Sci.\  {\bf 34} (1948) 211.

\bibitem{Buchb}I.~L.~Buchbinder and S.~M.~Kuzenko, {\em Ideas and methods of
supersymmetry and supergravity: or a walk through superspace}
(Institute of Physics Publishing, 1998).

\bibitem{Corson} E. M. Corson, {\it Introduction to tensors,
spinors and relativistic wave equations} (Blackie \& Son, 1953).

\bibitem{Bracken:1982ny}
A.~J.~Bracken and B.~Jessup,
J.\ Math.\ Phys.\  {\bf 23} (1982) 1925.

\bibitem{LV}
M.~A.~Vasiliev,
Fortsch.\ Phys.\  {\bf 35} (1987) 741;
V.~E.~Lopatin and M.~A.~Vasiliev,
Mod.\ Phys.\ Lett.\ A {\bf 3} (1988) 257.

\bibitem{Siegel:1986zi}
W.~Siegel and B.~Zwiebach,
Nucl.\ Phys.\ B {\bf 282} (1987) 125.

\bibitem{BB1}
X.~Bekaert and N.~Boulanger,
Commun.\ Math.\ Phys.\  {\bf 245} (2004) 27, {\tt hep-th/0208058}.

\bibitem{Pais}
A.~Pais and G.~E.~Uhlenbeck,
Phys.\ Rev.\  {\bf 79} (1950) 145.

\bibitem{F}
C.~Fronsdal,
Phys.\ Rev.\ D {\bf 18}, 3624 (1978);
\bibitem{FF}
J.~Fang and C.~Fronsdal,
Phys.\ Rev.\ D {\bf 18}, 3630 (1978).

\bibitem{fields}
W. Siegel, {\it Fields}, {\tt hep-th/9912205}.

\bibitem{d}
D.~Sorokin, {\it Introduction to the classical theory of higher
spins}, {\tt hep-th/0405069};\\
N.~Bouatta, G.~Compere and A.~Sagnotti, {\it An introduction to
free higher-spin fields,} {\tt hep-th/0409068};\\
M.~Bianchi, {\it Higher spins and stringy AdS(5) $\times$ S(5),} {\tt
hep-th/0409304}.
\bibitem{Francia:2002aa}
D.~Francia and A.~Sagnotti,
Phys.\ Lett.\ B {\bf 543} (2002) 303, {\tt hep-th/0207002}.

\bibitem{dMH}
P.~de Medeiros and C.~Hull,
Commun.\ Math.\ Phys.\  {\bf 235} (2003) 255, {\tt
hep-th/0208155}.

\bibitem{Francia:2002pt}
D.~Francia and A.~Sagnotti,
Class.\ Quant.\ Grav.\  {\bf 20} (2003) S473, {\tt
hep-th/0212185}.

\bibitem{Sagnotti:2003qa}
A.~Sagnotti and M.~Tsulaia,
Nucl.\ Phys.\ B {\bf 682} (2004) 83, {\tt hep-th/0311257}.

\bibitem{Pashnev:1998ti}
A.~Pashnev and M.~Tsulaia,
Mod.\ Phys.\ Lett.\ A {\bf 13} (1998) 1853
{\tt hep-th/9803207}.
\\
I.~L.~Buchbinder, V.~A.~Krykhtin and A.~Pashnev, {\it BRST
approach to Lagrangian construction for fermionic massless higher
spin fields,} {\tt hep-th/0410215}.


\bibitem{Olver:1983}
P.J.~Olver, {\it Differential hyperforms {I},} Univ. of Minnesota
report 82-101.

\bibitem{DuboisV} M.~Dubois-Violette and M.~Henneaux, {\em Lett. Math. Phys.}
{\bf 49} (1999) 245, {{\tt math.qa/9907135}};
Commun.\ Math.\ Phys.\  {\bf 226} (2002) 393, {\tt
math.qa/0110088}.

\bibitem{DD}
T.~Damour and S.~Deser,
Annales Poincare Phys.\ Theor.\  {\bf 47}, 277 (1987).

\bibitem{BB2}
X.~Bekaert and N.~Boulanger,
Phys.\ Lett.\ B {\bf 561}, 183 (2003) {\tt hep-th/0301243}.



\bibitem{BB3}
X.~Bekaert and N.~Boulanger, {\it Mixed symmetry gauge fields in a
flat background,}  {\tt hep-th/0310209}.


\bibitem{BBC}
X.~Bekaert, N.~Boulanger and S.~Cnockaert,
J. Math. Phys. {\bf 46} (2005) 012303, {\tt hep-th/0407102}.

\bibitem{Labastida:1986ft}
J.~M.~F.~Labastida,
Phys.\ Rev.\ Lett.\  {\bf 58} (1987) 531.

\bibitem{Labastida:1986zb}
J.~M.~F.~Labastida,
Phys.\ Lett.\ B {\bf 186} (1987) 365.



\bibitem{Hull:2001iu}
C.~M.~Hull,
JHEP {\bf 0109} (2001) 027, {\tt hep-th/0107149}.

\bibitem{Boulanger:2003vs}
N.~Boulanger, S.~Cnockaert and M.~Henneaux,
JHEP {\bf 0306} (2003) 060, {\tt hep-th/0306023};
\\
A.~S.~Matveev and M.~A.~Vasiliev, {\it On dual formulation for higher
spin gauge fields in (A)dS(d),} {\tt hep-th/0410249};
\\
K.~M.~Ajith, E.~Harikumar and M.~Sivakumar,
{\it Dual linearised gravity in arbitrary dimensions from Buscher's
construction,}
{\tt hep-th/0411202}.


\bibitem{Burdik:2001hj}
C.~Burdik, A.~Pashnev and M.~Tsulaia,
Mod.\ Phys.\ Lett.\ A {\bf 16} (2001) 731, {\tt hep-th/0101201};
Nucl.\ Phys.\ Proc.\ Suppl.\  {\bf 102} (2001) 285, {\tt
hep-th/0103143}.


\bibitem{Bonelli:2003kh}
G.~Bonelli,
Nucl.\ Phys.\ B {\bf 669} (2003) 159, {\tt hep-th/0305155}.

\bibitem{Bekaert:2003uc}
X.~Bekaert, I.~L.~Buchbinder, A.~Pashnev and M.~Tsulaia,
Class.\ Quant.\ Grav.\  {\bf 21} (2004) S1457, {\tt
hep-th/0312252}.

\bibitem{Barnich:2004cr}
G.~Barnich, M.~Grigoriev, A.~Semikhatov and I.~Tipunin, {\it
Parent field theory and unfolding in BRST first-quantized terms,}
{\tt hep-th/0406192}.

\bibitem{Brink:2000ag}
L.~Brink, R.~R.~Metsaev and M.~A.~Vasiliev,
Nucl.\ Phys.\ B {\bf 586} (2000) 183 {\tt hep-th/0005136};\\
Y.~M.~Zinoviev,
{\it On massive mixed symmetry tensor fields in Minkowski space and (A)dS,}
{\tt hep-th/0211233};\\
K.~B.~Alkalaev, O.~V.~Shaynkman and M.~A.~Vasiliev,
Nucl.\ Phys.\ B {\bf 692} (2004) 363 {\tt hep-th/0311164};
\\
P.~de Medeiros,
Class.\ Quant.\ Grav.\  {\bf 21} (2004) 2571
{\tt hep-th/0311254};
\\
R.~R.~Metsaev,
{\it Mixed symmetry massive fields in AdS(5),}
{\tt hep-th/0412311}.

\bibitem{dw}
S.~Deser and A.~Waldron,
Nucl.\ Phys.\ B {\bf 607} (2001) 577 {\tt hep-th/0103198}.
\\
S.~Deser and A.~Waldron, {\it Conformal invariance of partially
massless higher spins,} {\tt hep-th/0408155}.

\bibitem{Siegel:1988gd}
G.~Mack and A.~Salam,
Annals Phys.\  {\bf 53} (1969) 174;\\
E.~Angelopoulos, M.~Flato, C.~Fronsdal and D.~Sternheimer,
Phys.\ Rev.\ D {\bf 23} (1981) 1278.
\\
W.~Siegel,
Int.\ J.\ Mod.\ Phys.\ A {\bf 4} (1989) 2015;\\
E.~Angelopoulos and M.~Laoues,
Rev.\ Math.\ Phys.\  {\bf 10} (1998) 271 {\tt hep-th/9806100};\\
R.~R.~Metsaev,
Mod.\ Phys.\ Lett.\ A {\bf 10} (1995) 1719;\\
S. Ferrara and C. Fronsdal, {\it Conformal fields in higher
dimensions}, {\tt hep-th/0006009}.

\bibitem{cd3}
E.~S.~Fradkin and V.~Y.~Linetsky,
Mod.\ Phys.\ Lett.\ A {\bf 4} (1989) 731 [Annals Phys.\  {\bf 198}
(1990) 293].
\\
C.~N.~Pope and P.~K.~Townsend,
Phys.\ Lett.\ B {\bf 225} (1989) 245.


\bibitem{Shaynkman:2004vu}
O.~V.~Shaynkman, I.~Y.~Tipunin and M.~A.~Vasiliev, {\it Unfolded
form of conformal equations in M dimensions and o(M+2)-modules,}
{\tt hep-th/0401086}.
\bibitem{Hull:2000rr}
C.~M.~Hull,
Nucl.\ Phys.\ B {\bf 583} (2000) 237, {\tt hep-th/0004195};
Class.\ Quant.\ Grav.\  {\bf 18} (2001) 3233, {\tt
hep-th/0011171};
JHEP {\bf 0012} (2000) 007, {\tt hep-th/0011215}.


\bibitem{Bandos:2001pu}
I.~A.~Bandos, J.~A.~de Azc\'arraga, J.~M.~Izquierdo and
J.~Lukierski,
Phys.\ Rev.\ Lett.\  {\bf 86} (2001) 4451, {\tt hep-th/0101113}.

\bibitem{Martin}
M. Cederwall, {\it AdS Twistors for Higher Spin Theory},
hep-th/0412222.

\bibitem{Kugo}
T.~Kugo and P.~K.~Townsend,
Nucl.\ Phys.\ B {\bf 221} (1983) 357.
\\
P. S. Howe, G. Sierra and  P. K. Townsend, {Nucl. Phys.} {\bf B221}
(1983) 331.
\\
P. West, {\em Supergravity, Brane Dynamics and String Duality},
hep-th/9811101.
\bibitem{bem}
I. Bengtsson and M. Cederwall, Nucl. Phys. B {\bf 302} (1988) 81.
\bibitem{BHS}
X.~Bekaert, M.~Henneaux, and A.~Sevrin, Commun. Math. Phys. {\bf
224} (2001) 683, {\tt hep-th/0004049}.


\bibitem{Bandos:2004nn}
I.~Bandos, P.~Pasti, D.~Sorokin and M.~Tonin,
JHEP {\bf 0411} (2004) 023 {\tt hep-th/0407180}.



\bibitem{KennedyVanProey}
A.~D.~Kennedy,
J.\ Math.\ Phys.\  {\bf 22}, 1330 (1981);\\
J.~W.~van Holten and A.~Van Proeyen,
J.\ Phys.\ A {\bf 15} (1982) 3763;\\
A.~Van Proeyen, {\it Tools for supersymmetry,} {\tt
hep-th/9910030}.






\end{thebibliography}
\end{document}